\documentclass[10pt,journal,compsoc]{IEEEtran}
\usepackage{graphicx,epstopdf}
\usepackage[subrefformat=parens,labelformat=parens]{subfig}
\usepackage{adjustbox}
\usepackage[table,xcdraw]{xcolor}
\usepackage{hyperref}

\ifCLASSOPTIONcompsoc
  \usepackage[nocompress]{cite}
\else
  \usepackage{cite}
\fi
\ifCLASSINFOpdf
\else
\fi
\usepackage[cmex10]{amsmath}
\usepackage{mathtools}

\usepackage{siunitx}
\usepackage[numbers,sort&compress]{natbib}
\newcommand{\equationname}{Eq.}

\usepackage{mystyle}
\usepackage{algorithm}
\usepackage{algpseudocode}
\usepackage{multirow}
\makeatletter
\def\BState{\State\hskip-\ALG@thistlm}
\makeatother
\usepackage{stfloats}
\usepackage{xcolor}
\usepackage[super]{nth}

\usepackage{comment}
\usepackage[bottom]{footmisc}
\def\baselinestretch{0.85}

\begin{document}
%
\title{RAT: Reinforcement-Learning-Driven and  Adaptive Testing for Vulnerability Discovery in Web Application Firewalls\\
\thanks{© 2021 IEEE.  Personal use of this material is permitted.  Permission from IEEE must be obtained for all other uses, in any current or future media, including reprinting/republishing this material for advertising or promotional purposes, creating new collective works, for resale or redistribution to servers or lists, or reuse of any copyrighted component of this work in other works.}}
%
%
%
%

\author{Mohammadhossein~Amouei,
        Mohsen~Rezvani,
        Mansoor~Fateh
\IEEEcompsocitemizethanks{\IEEEcompsocthanksitem M. Amouei, M. Rezvani and M. Fateh are with the Faculty of Computer Engineering, Shahrood University of Technology, Iran.\protect\\
E-mail: \{mhamooei,mrezvani,mansoor\_fateh\}@shahroodut.ac.ir
}
}

%
%

\markboth{IEEE Transactions on Dependable and Secure Computing}%
{This article has been accepted for publication in IEEE Transactions on Dependable and Secure Computing. This is the author's version which has not been fully edited and content may change prior to final publication. Citation information: DOI 10.1109/TDSC.2021.3095417}
%



\IEEEtitleabstractindextext{%
\begin{abstract}
Due to the increasing sophistication of web attacks, Web Application Firewalls (WAFs) have to be tested and updated regularly to resist the relentless flow of web attacks. In practice, using a brute-force attack to discover vulnerabilities is infeasible due to the wide variety of attack patterns. Thus, various black-box testing techniques have been proposed in the literature. However, these techniques suffer from low efficiency. This paper presents Reinforcement-Learning-Driven and Adaptive Testing (\textit{RAT}), an automated black-box testing strategy to discover injection vulnerabilities in WAFs. In particular, we focus on SQL injection and Cross-site Scripting, which have been among the top ten vulnerabilities over the past decade. More specifically, \textit{RAT} clusters similar attack samples together. It then utilizes a reinforcement learning technique combined with a novel adaptive search algorithm to discover almost all bypassing attack patterns efficiently. We compare \textit{RAT} with three state-of-the-art methods considering their objectives. The experiments show that \textit{RAT} performs 33.53\% and 63.16\% on average better than its counterparts in discovering the most possible bypassing payloads and reducing the number of attempts before finding the first bypassing payload when testing well-configured WAFs, respectively.
\end{abstract}

\begin{IEEEkeywords}
Security testing, injection attack, adaptive testing, web application firewall (WAF), test case clustering.
\end{IEEEkeywords}}

\maketitle

\IEEEdisplaynontitleabstractindextext

%
\IEEEpeerreviewmaketitle

\IEEEraisesectionheading{\section{Introduction}\label{sec:introduction}}

%
%
%
%
\IEEEPARstart{I}{n} recent decades, most traditional brick and mortar businesses have transformed into online ones, such as online shopping, e-banking, social media, etc. Thus, an enormous amount of private data of individuals and organizations is stored in web applications databases, making them tempting targets for attackers. A recent report reveals that web applications may experience up to 26 attacks per minute \cite{simos2019practical}. Moreover, according to Symantec's security report, 76\% of websites are vulnerable to several attacks \cite{Chandrasekar2017}. 

A proper way to provide the security is to use Web Application Firewalls (WAFs) which analyze HTTP(S) traffic to prevent malicious requests from reaching the web applications. The Open Web Application Security Project (OWASP\footnote{\url{https://owasp.org}}) defines WAF as \lq{}\lq{}a security solution on the web application level which - from a technical point of view - does not depend on the application itself.\rq{}\rq{} \cite{chapter2008owasp}. To put it in perspective, WAFs intercept bi-directional HTTP(S) traffic, analyze it, and decide whether it is malicious or benign. Common rule-based WAFs use a set of rules to make a decision. For instance, WAFs utilize regular expressions to detect SQL injection (SQLi) attacks. Moreover, recently, extensive research has been done on intelligent Machine-Learning-Based (ML-Based) WAFs to distinguish between malicious and benign traffic with machine learning algorithms \cite{tekerek2019design,vartouni2019leveraging,mac2018detecting}.

\begin{figure}[!t]
\centering
  \includegraphics[width=\linewidth]{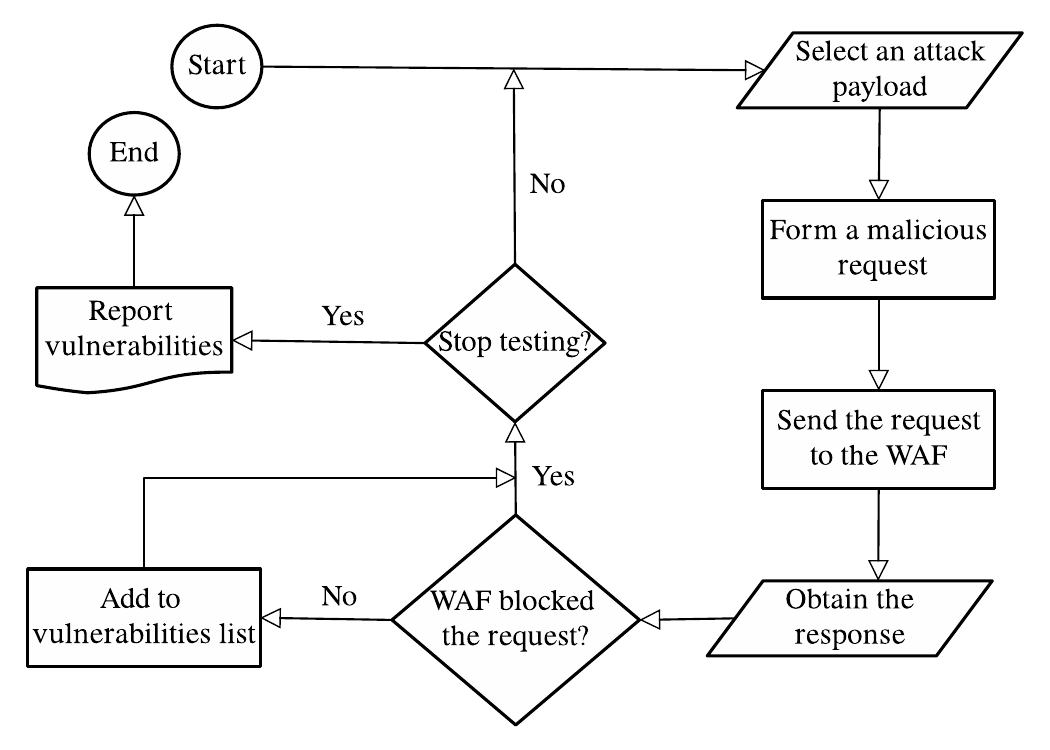}
  \caption{Testing procedure for a WAF to discover vulnerabilities.}
  \label{fig:test}
 \vspace{-2em}
\end{figure}

A recent study confirmed that web attacks had grown in size and sophistication \cite{zhang2019art4sqli}; consequently, it is crucial to regularly test and maintain WAFs to keep them secure and efficient. In \figurename~\ref{fig:test}, the procedure for testing a WAF is illustrated. As web attacks become more sophisticated, traditional WAF rules become more complex, and ML-Based WAFs tend to learn novel attacks. At the same time, manual testing and maintenance become more arduous tasks. Testing WAFs is also extremely expensive, especially in time, due to the massive variety of attacks. Hence, optimal automated testing is essential for WAFs to protect web applications and services efficiently.

One of the common and destructive categories of attacks is injection. Injections are attacks in which the attacker injects a malicious input to a web application. Then, the application interprets this input as a part of a command or a query, which can result in severe damages. In this paper, we focus our tests on two common injection attacks: SQLi and Cross-site Scripting (XSS). These attacks are reported as top 10 vulnerabilities \cite{wichers2017owasp} and have attracted a lot of attention\cite{simos2019practical,bozic2015attack,simos2016combinatorial,simos2019automated, thome2014search, avancini2011security, duchene2012xss, demetrio2020waf, tripp2013finding, appelt2015behind, appelt2018machine, zhang2019art4sqli, lv2019adaptive}.

There are various types of security testing proposed in the literature, such as white-box testing, model-based testing, and black-box testing. However, these techniques suffer from limitations that can affect their practical applicability. White-box testing needs access to the applications' source code, which might not be possible in testing industrial applications. Moreover, each white-box testing tool supports only some specific programming languages; thus, they can only test applications developed with those programming languages. Model-based testing techniques require a model representing the security policies, which is difficult to create and is often unavailable. The black-box testing, nevertheless, does not have the mentioned limitations of the two other techniques \cite{appelt2018machine}. However, despite the remarkable effort that has been devoted to black-box testing, studies indicate that black-box testing is inefficient, and a large number of vulnerabilities remains undiscovered \cite{mcgraw2004software,khan2012comparative}.

In recent years, with the power of artificial intelligence, novel black-box testing techniques have shown a significant improvement in efficiency and effectiveness \cite{tripp2013finding,appelt2015behind,appelt2018machine}. These techniques utilize artificial intelligence methods, such as evolutionary algorithms \cite{thome2014search, avancini2011security, duchene2012xss} or machine learning techniques \cite{elderman2016adversarial, demetrio2020waf, appelt2015behind} to improve the black-box testing performance. However, these techniques still consume many requests, and yet many vulnerabilities remain undetected. Therefore, black-box testing requires further improvement.

This research aims to design a practical automated black-box security testing approach to uncover WAFs' vulnerabilities efficiently. Thus, in this paper, we propose a method called \textit{RAT}, which uses machine learning algorithms to provide better effectiveness and efficiency to the black-box testing. \textit{RAT} first tokenizes attack payloads using $n$-gram. It then clusters similar attack payloads and uses a reinforcement learning technique called decayed $\epsilon$-greedy policy combined with a novel adaptive search technique to find its way through the testing jungle.

Furthermore, we compare \textit{RAT} with three state-of-the-art techniques, including \textit{Ml-Driven E} \cite{appelt2018machine}, \textit{ART4SQLi} \cite{zhang2019art4sqli} and \textit{XSSART} \cite{lv2019adaptive}. These tests are designed considering our counterparts' objectives. More specifically, \textit{Ml-Driven E} tend to discover the highest possible number of SQLi vulnerabilities, and \textit{ART4SQLi} and \textit{XSSART} aim to find the very first bypassing payload with the lowest number of requests. Thus, to compare \textit{RAT}  with \textit{Ml-Driven E} we measure the total bypassing payloads within a limited number of requests for both techniques, and for the comparison between \textit{RAT} and \textit{ART4SQLi} and \textit{XSSART}, we compare the number of blocked payloads before finding the first bypassing payload for all three techniques. We use SQLi dataset to compare \textit{RAT} with \textit{Ml-Driven E} and \textit{ART4SQLi}, and XSS dataset to compare \textit{RAT} with \textit{XSSART}. We also compare \textit{RAT} with a simple random testing technique (we name it \textit{Random Fuzzer}) as a basic method. Our comparative experiments show that \textit{RAT} can discover an average of 33.53\% more bypassing attack than \textit{Ml-Driven E} within a limited number of requests. Moreover, according to our experiments, our adaptive search algorithm is an average of 61.43\% faster than \textit{ART4SQLi} when facing well-configured WAFs. However, \textit{ART4SQLi} could discover the first bypassing payload about 38.70\% faster than \textit{RAT} in testing a WAF with massive vulnerabilities. Moreover, results show that \textit{RAT} is an average of 64.90\% faster than \textit{XSSART} in finding the first bypassing payload.

The main contributions of the paper summarized as:
\begin{enumerate}
\item Since string-based injection attack payloads are sequences of specific string tokens, we employ $n$-gram, known for simplicity and scalability \citep{mnih2009scalable}, as a feature extraction method. In our experiments, we observed that $n$-gram could model sophisticated attack patterns while extracting significantly fewer features than the \textit{ML-Driven E}, resulting in a better efficiency than \textit{ML-Driven E}.
\item We evaluate the effects of the clustering and propose a method to cluster the similar attack payloads.
\item We use the $\epsilon$-greedy algorithm and a novel adaptive search technique to enhance the efficiency of our black-box testing approach.
\end{enumerate}

The remaining sections are organized as follows. Section~\ref{related_work} describes SQLi and XSS attacks as well as previous research. Section~\ref{approach} details the proposed approach. In Section~\ref{empirical}, we explain research questions as well as the experimental environment. Section~\ref{results} discusses the experiments and evaluation results. Section~\ref{conclusion} concludes this paper.
\vspace{-1.5em}
\section{Background and Related Work} \label{related_work}
This section aims to provide a brief description of the code injection attacks, such as SQL Injection and Cross-Site Scripting. In the final subsection, we narrate the research story of black-box testing for these two web attacks.
\vspace{-1em}
\subsection{SQL Injection}
SQL databases, also known as relational, are the most popular ones among developers \cite{overflow2019stack}. Web-based applications that use SQL databases communicate with database engines by statements that are defined in a language, named SQL. These statements are usually formed dynamically by concatenating various substrings. Each substring is provided either by a user or the application itself. Once statement formation has done, the database engine executes it.

\begin{figure}[!t]
\centering
\begin{lstlisting}[style=JavaStyle]
 public void doPost(HttpServletRequest request, HttpServletResponse response)
 throws ServletException, IOException {

  String user_name = request.getParameter("username");
  String user_pass = request.getParameter("password");
  
  String sql_statement = "SELECT * FROM users WHERE username = '" 
  + user_name 
  + "' AND password = '"
  + user_pass + "' ";
 
 result = Database.execute(sql_statement)
}
\end{lstlisting}
\caption{Example of an unsafe SQL statement formation in Java.}
\label{fig:sql_ex}
\vspace{-2em}
\end{figure}

\figurename~\ref{fig:sql_ex} shows an instance of an unsafe SQL statement formation. In \figurename~\ref{fig:sql_ex}, the variables \texttt{user\_name} and \texttt{user\_pass} are provided by the user through HTTP request (line 4-5). Then, they are concatenated with the SQL statement without any sanitization and stored in the variable \texttt{sql\_statement} (line 7-10). Finally, the \texttt{sql\_statement} passed to the function \texttt{execute} of the class \texttt{Database} to be executed by the database server (line 12). In case of malicious inputs, database server executes the infected statement and performs the attacker's desired action.

SQLi is an injection attack in which the attacker targets applications with unsafe SQL statement formation, and passes malicious input parameters to the victim application and misleads the application to perform an unauthorized action (e.g., accessing confidential data without granting required permission). For instance, in \figurename~\ref{fig:sql_ex}, if instead of passing a real username to the variable \texttt{user\_name}, an attacker fills the username parameter with \texttt{` OR 1 = 1 \#} and leaves the password empty, then the value of the \texttt{sql\_statement} would be \texttt{SELECT * FROM users WHERE username = `` OR 1 = 1 \# AND password = ``} which is equal to \texttt{SELECT * FROM users WHERE username = `` OR 1 = 1}. In SQL, the symbol \texttt{\#} is an inline comment operator, and \texttt{1 = 1} is an always True condition, thus if the database executes the resulting statement, it returns all rows of the table \texttt{users}.

SQLi attacks are a high-risk security threat to organizations. By a successful SQLi exploit, confidential data of organizations or individuals can be modified or stolen. Thus,  protecting applications against SQLi attacks is vital.
\vspace{-1em}
\subsection{Cross-Site Scripting (XSS)}
XSS is a code injection attack in which the attacker injects malicious scripts into legitimate websites to execute them in the web browser of end-users. Any web application that generates output using input from users can be vulnerable to such an attack. XSS occurs when a victim visits an infected web application that carries the malicious script to the browser. Within the browser, the script can access user's sensitive information (e.g., stealing costumer's payment info), cookies and, any other data that related to the web application and retained by the user's browser.

\begin{figure}[!t]
\centering
\includegraphics[width=\linewidth]{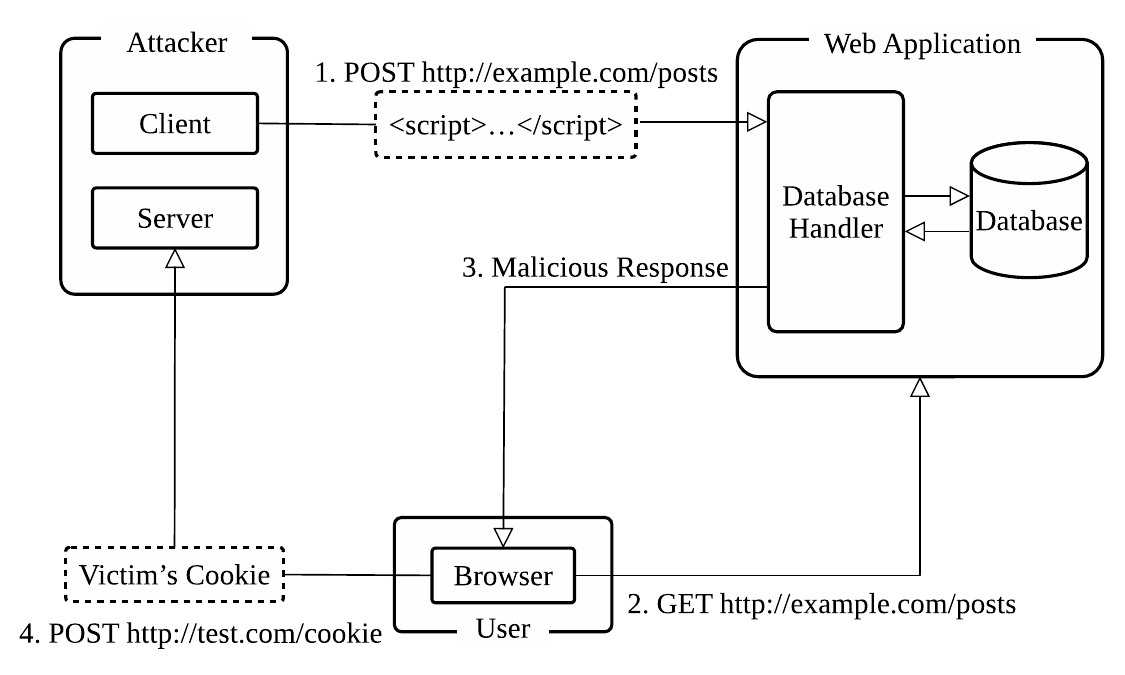}
\caption{Example of an XSS attack attempt, in which the attacker steals the victim's cookie by submitting a malicious post to a vulnerable social media.}
\label{fig:xss_ex}
\vspace{-1em}
\end{figure}

\begin{figure}[!t]
\centering
\begin{lstlisting}[style=htmlcssjs]
<!DOCTYPE html>
<html>
<h1> Latest Posts </h1>
...
<script>
  $(document).ready(function() {
      var cookie = document.cookie;
      $.post("http://test.com/cookie", 
      {cookie: cookie},
      function(result) {});
  });
</script>
...
</html>
\end{lstlisting} 
\caption{Example of a malicious HTML containing JavaScript to steal user's cookie.}
\label{fig:malicious_script}
\vspace{-2em}
\end{figure}

\figurename~\ref{fig:xss_ex} shows an instance of an XSS attack attempt in which the attacker tries to steal the user's cookie of a vulnerable social media. The details are as follows:
\begin{enumerate}
	\item The attacker submits a malicious post containing JavaScript code to the application, which inserts it into the application's database without sanitization.
	\item The victim user sends a GET request to the application to get the latest posts. Afterward, the application retrieves the posts, including the malicious one, from its database and locates them into the HTML page as the response to the client.
	\item The application responds to the user with a page containing the malicious script (Figure \ref{fig:malicious_script}).
	\item Once the user's browser receives the page, render it and executes its scripts, and as a result, the JavaScript code in Figure \ref{fig:malicious_script} (line 5-12) sends the user's cookie to the attacker's server.
\end{enumerate}

\vspace{-2em}
\subsection{Related Work}
Over the past decade, both SQLi and XSS attacks have been attractive topics for researchers, and remarkable research efforts have been made toward vulnerability detection methods, particularly automated black-box testing \citep{felderer2016security}.

Various combinatorial testing methods have been proposed in the literature for both XSS and SQLi attacks \cite{simos2019practical,bozic2015attack,simos2016combinatorial,simos2019automated}. These methods parametrize attack patterns in the form of BNF grammar rules. Then they test combinations of parameters by covering \textit{t-way} interactions in which $t$ is the number of parameters, and a higher value for $t$ can produce more complex patterns. However, a large $t$ consumes a vast number of HTTP requests.

Knowing the capability of Artificial Intelligence (AI), researchers tend to provide better-optimized solutions using AI for various problems, including security assessment. For example, \citeauthor{thome2014search} \citep{thome2014search} proposed a fitness function to measure how close an SQLi literal is from generating a bypassing payload. \citeauthor{avancini2011security} \citep{avancini2011security} used the Genetic Algorithm (GA) to find vulnerable inputs in webpage. Moreover, \citeauthor{duchene2012xss} \citep{duchene2012xss} applied GA to generate XSS payloads in fuzz testing. There are also adversarial and learning-based methods which we describe in the following.
\subsubsection{Adversarial}
\citeauthor{elderman2016adversarial} \citep{elderman2016adversarial} simulated an adversarial cybersecurity game in which an attacker and a defender are two adversarial agents that use reinforcement learning techniques to win the game. \citeauthor{demetrio2020waf} \citep{demetrio2020waf} proposed \textit{WAF-A-MoLE}, an adversarial method for mutating attack strings to bypass ML-Based WAFs. Mostly, adversarial methods aim to bypass ML-Based WAFs, whereas our method targets signature-based WAFs. Nevertheless, our approach can be combined with adversarial techniques to achieve higher performance. For instance, it can be integrated into \textit{WAF-A-MoLE} to function as a guide to increase its efficiency.

\subsubsection{Learning-based}
\citeauthor{tripp2013finding} \citep{tripp2013finding} suggested \textit{XSS Analyzer}, a web security testing approach with learning capability. The authors proposed a method that learns from previous attempts to select the next payload with a higher probability of exposing a vulnerability in the System Under Test (SUT). In particular, \textit{XSS Analyzer} generates payloads for a grammar, devised based on a comprehensive dataset of XSS payloads, and learns which tokens are preventing an attack from evading the security.

Inspired by \textit{XSS Analyzer}, \citeauthor{appelt2015behind} \citep{appelt2015behind,appelt2018machine} proposed \textit{ML-Driven B}, \textit{ML-Driven D} and later on, \textit{ML-Driven E}, learning approaches to SQLi vulnerability detection. The same as \textit{XSS Analyzer}, \textit{ML-Drivens} benefit from the idea of learning literals that prevent an attack from bypassing the security from previous attempts. The main difference is within the learning process. Whereas \textit{XSS Analyzer} only learns individual literals, \textit{ML-Drivens} learn the combinations of literals. More specifically, \textit{ML-Drivens} split each attack's derivation tree into subtrees and then measures the likelihood of bypassing attacks using a decision tree. The authors prove that \textit{ML-Drivens} can learn more complex attack patterns and outperform state-of-the-art tools, thus, suggesting that learning combinations of literals can effectively guide the testing procedure.

Usually, bypassing payloads are rare in a comprehensive payload collection which makes it challenging to find effective payloads with a reasonable number of tests. \citeauthor{zhang2019art4sqli} \citep{zhang2019art4sqli} proposed \textit{ART4SQLi}, an adaptive random testing approach for SQLi vulnerability and \citeauthor{lv2019adaptive} \citep{lv2019adaptive} proposed \textit{XSSART} an adaptive random testing approach for XSS vulnerability detection. Both \textit{ART4SQLi} and \textit{XSSART} use similarity metrics to expedite the process of finding an effective attack payload, and demonstrate that payload spaces are sparse, and bypassing payloads tend to cluster together. Thus, in this research, we consider clustering similar payloads and evaluate the effects of clustering on efficiency.

Although \textit{ML-Drivens} can discover a large number of bypassing payloads, they require a large number of observations to learn effective patterns. \textit{ML-Drivens} require an initial collection of both passed and blocked attacks to train a decision tree; thus, before training the decision tree, they perform a random search which, facing a well-configured WAF, consumes a large number of requests. In contrast to \textit{ML-Drivens}, our adaptive search technique only uses blocked attacks to uncover bypassing ones in the very beginning of the test. It then uses the discovered bypassing attacks to improve its performance.

Moreover, \textit{ML-Drivens} break payload derivation trees into sub-trees to use them as features, creating a vast feature space that exponentially expands if we try to test longer attack payloads (e.g., XSS). Training a decision tree with this massive number of features is non-practical; thus, the authors suggested selecting a small random subset of the feature space, resulting in low efficiency of their approach. In comparison, RAT showed a better performance in our experiments, since first of all, n-gram extracts significantly fewer features than \textit{ML-Drivens} feature extraction algorithm. Secondly, our clustering and feature reduction technique can effectively reduce the number of features, resulting in better effectiveness and efficiency.

Last but not least, the decision tree algorithm suffers from the local minima problem, and it is unstable as small changes in training data remarkably affect classification results \citep{witten2005practical}. Thus, the authors in \citep{appelt2018machine} discussed using an ensemble model such as a random forest algorithm may address these problems. However, it adds high computational overhead as multiple decision trees have to be trained during each update, and their experiments showed that its improvement is not significant enough. Similar to \textit{ML-Drivens}, in our work, the $\epsilon$-greedy policy suffers from local optima, and it is unstable because the cluster selection order can significantly change the final results. Nevertheless, we try to avoid local optima by controlling exploration and exploitation rate using decayed $\epsilon$-greedy policy, which does not add any computational burden. More specifically, this policy starts with a big $\epsilon$ to explore and find clusters that include bypassing attacks. Since our adaptive search algorithm can quickly find bypassing payloads inside clusters, decayed $\epsilon$-greedy can quickly filter the clusters with non-bypassing payloads. Then, the $\epsilon$ decreases, so the \textit{RAT} focuses on the clusters with bypassing attacks. These quick exploration and exploitation significantly improve \textit{RAT}'s stability and efficiency. Our experiments show that the \textit{RAT} is considerably more stable and efficient than \textit{ML-Driven E}.
\vspace{-1.5em}
\section{Approach} \label{approach}
In this section, we first describe an overview of the proposed approach. We then explain the details of each module in next subsections.

\begin{figure}[!t]
\centering
  \includegraphics[width=\linewidth]{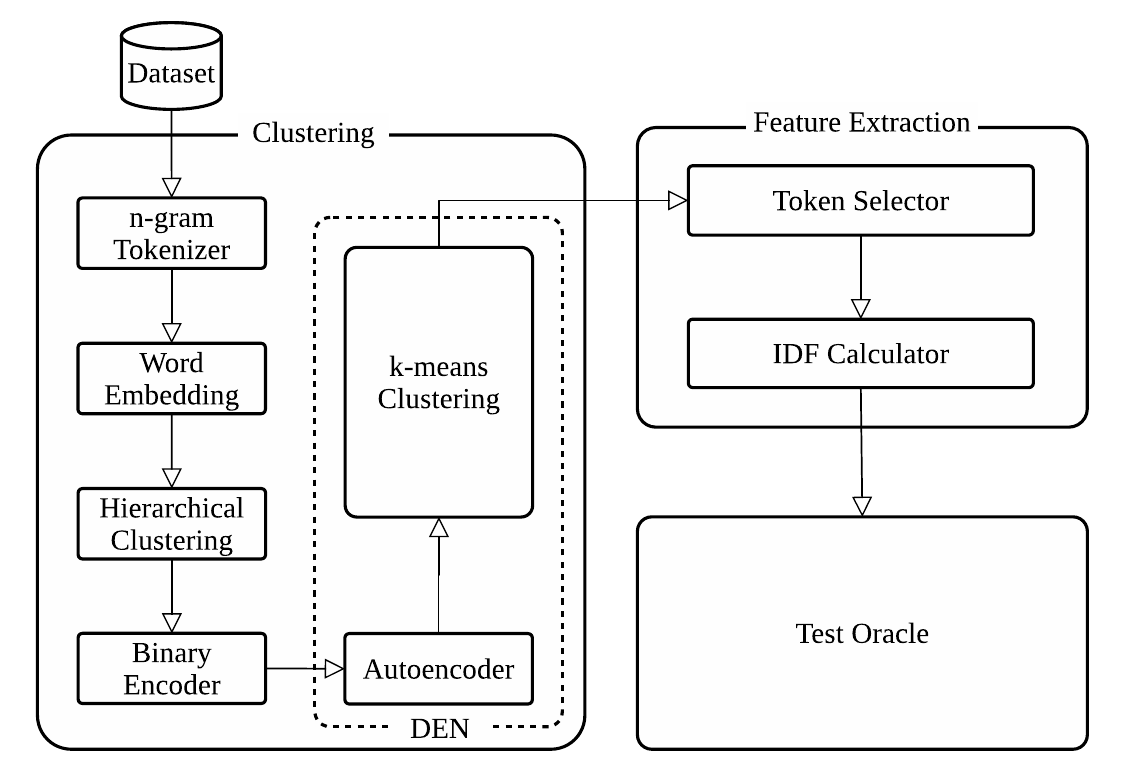}
  \caption{Overview of the first phase of the proposed approach, which shows the preparation of the test oracle.}
  \label{fig:overview}
  \vspace{-2em}
\end{figure}

\vspace{-1.5em}
\subsection{Framework Overview} \label{basicidea}
The code injection attack payload is formed by concatenating miscellaneous string fragments. Considering these fragments are the test parameters, each fragment is responsible for either failure or success of an attack payload in circumventing the firewall. In black-box testing, we do not have any information about the source code of the application under the test to distinguish effective from ineffective fragments. Nevertheless, it is possible to find effective fragments based on feedback obtaining from the application during the test process. However, due to the large variety of fragments in a rich dataset, a testing tool needs too many observations to gather enough information. Therefore we consider clustering similar payloads together; thus, we can reduce the variety of fragments to distinguish between them rapidly. Moreover, since bypassing payloads tend to cluster together, devoting search effort to effective clusters significantly improves the performance.

\figurename~\ref{fig:overview} illustrates an overview of \textit{RAT}. It shows the process of preparing attack samples for the testing phase. Here, \texttt{Dataset} is the collection of attack payloads (see section \ref{sec-dataset} for more details about the datasets). At the very first step, \texttt{$n$-gram Tokenizer} (Section~\ref{n-gram}) tokenizes the attack samples of the dataset. Then, the \texttt{Word Embedding} (Section \ref{we}) module maps each token to the vector of real numbers, and \texttt{Hierarchical Clustering} (Section~\ref{hc}) clusters tokens using these vectors. \texttt{Binary Encoder} (Section~\ref{be}) then forms a binary feature vector using clusters that are obtained from \texttt{Hierarchical Clustering} module, and passes these vectors to \texttt{Deep Embedding Module (DEN)} (Section~\ref{den}) to cluster attack samples. Finally, in the feature extraction phase, \texttt{Token Selector} (Section~\ref{ts}) selects only effective tokens for each cluster, and then, \texttt{IDF Calculator} (Section~\ref{idfcalculator}) forms the main feature vector for each attack payload. Clusters, attack samples, and their feature vectors are then passed to the \texttt{Test Oracle} (Section~\ref{oracle}) to be used in our testing algorithm.
\vspace{-1em}
\subsection{Clustering Payloads} \label{clustering}
At the very first stage of our approach, we decompose attack payloads into string fragments and then cluster them together based on their common fragments. In other words, we split the dataset into smaller datasets; thus, later, we can search in each mini dataset independently. The following sections detail the process, and our experiments show that clustering significantly reduces the number of features. This feature reduction is important as a high number of features require many observations to learn patterns; thus, removing irrelevant features improves accuracy and efficiency - reducing features decreases the computational complexity and working with small feature vectors requires fewer resources (e.g. disk storage or memory) than big feature vectors \citep{singh2017document}.
\begin{figure}[!t]
\centering
  \includegraphics[width=\linewidth]{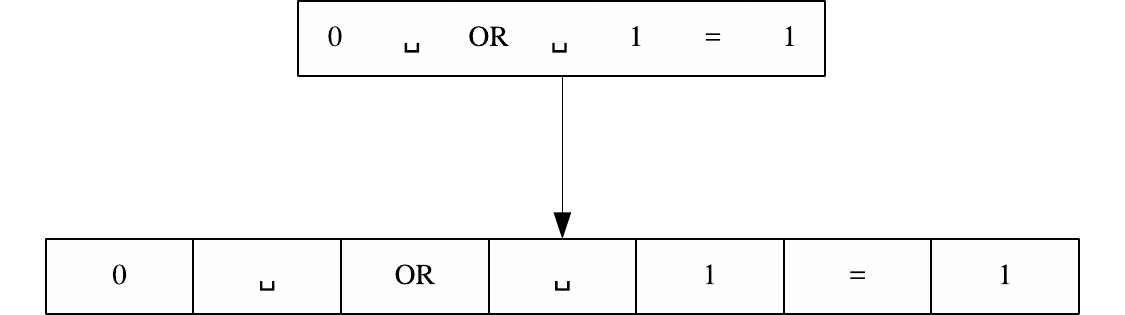}
  \caption{Decomposition of a sample attack payload to its constitutive tokens.}
  \label{fig:ngram}
  \vspace{-0.5em}
\end{figure}

\subsubsection{$n$-gram Tokenizer} \label{n-gram}
We can decompose every payload of a code injection attack into smaller pieces called tokens. Each token can be a single character or a sequence of characters. As an example, in SQLi, constitutive tokens of the sample payload showed in \figurename~\ref{fig:ngram}. Simply, each token is responsible for either failure or success of an attack payload; however, not always a single token but also a combination of tokens can lead a payload to success. Therefore, we extract combinations of tokens instead of single tokens using a well-known method called $n$-gram. In this paper, we refer to n-gram extracted tokens as fragments.

\begin{table}[!t]
\caption{Example of payload decomposition for the payload \texttt{0)\textvisiblespace or\textvisiblespace not\textvisiblespace  0\textgreater(\textvisiblespace !\texttildelow \textvisiblespace 0)--} using $n$-gram.}
\label{tab:n-gram}
\begin{adjustbox}{center}
\renewcommand{\arraystretch}{1.5}
\begin{tabular}{|c|c|c|}
\hline
\begin{tabular}[c]{@{}c@{}}Unigram\\ ($n=1$)\end{tabular} &
 \begin{tabular}[c]{@{}c@{}}Bigram\\ ($n=2$)\end{tabular} & 
 \begin{tabular}[c]{@{}c@{}}Trigram\\ ($n=3$)\end{tabular} \\ \hline
0 , ) , \textvisiblespace \space , ...&
 0) , )\textvisiblespace \space , \textvisiblespace or , ...& 
 0)\textvisiblespace \space , )\textvisiblespace or , \textvisiblespace or\textvisiblespace \space , ...                                           \\ \hline
\end{tabular}
\end{adjustbox}
\vspace{-1em}
\end{table}

$N$-gram is a contiguous sequence of $N$ tokens of a string. We use $n$-gram to decompose attack payloads into smaller fragments, which help us consider more complex patterns in our search strategy. Table~\ref{tab:n-gram} shows an example of payload decomposition for the payload \texttt{0)\textvisiblespace or\textvisiblespace not\textvisiblespace  0\textgreater(\textvisiblespace !\texttildelow \textvisiblespace 0)--} with three different values for $N$. As shown in Table~\ref{tab:n-gram}, with the bigger $N$, we can extract more complex patterns; however, as we increase the size of $N$, the variety of fragments grows. In our empirical study, we investigate the effect of $N$ on search results for each dataset.

\begin{figure}[!t]
\centering
  \includegraphics[width=\linewidth]{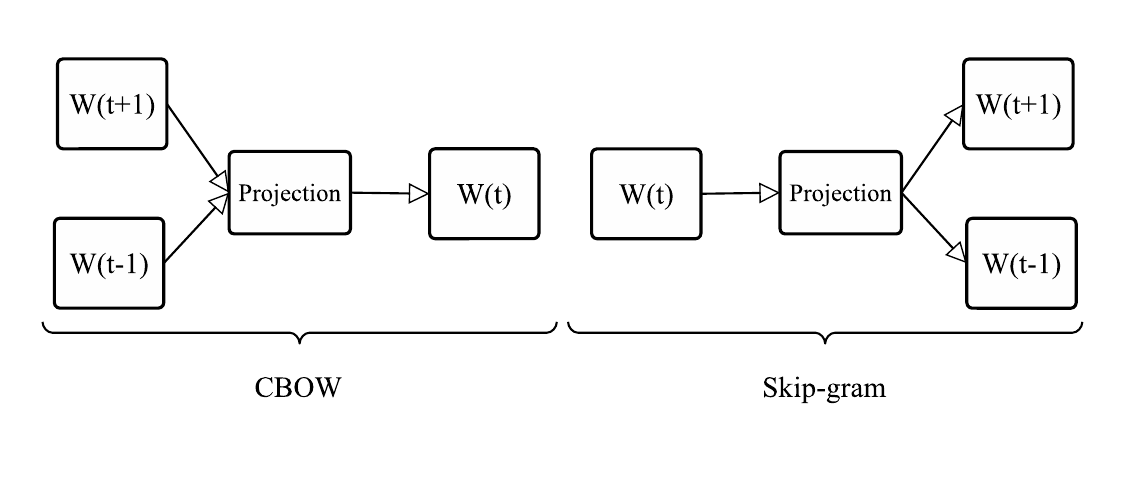}
  \caption{Architectures of Word2Vec models: Continuous Bag of Words (CBOW) and Skip-gram.}
  \label{fig:we_models}
  \vspace{-1em}
\end{figure}
\begin{figure}[!t]
\centering
  \includegraphics[width=\linewidth]{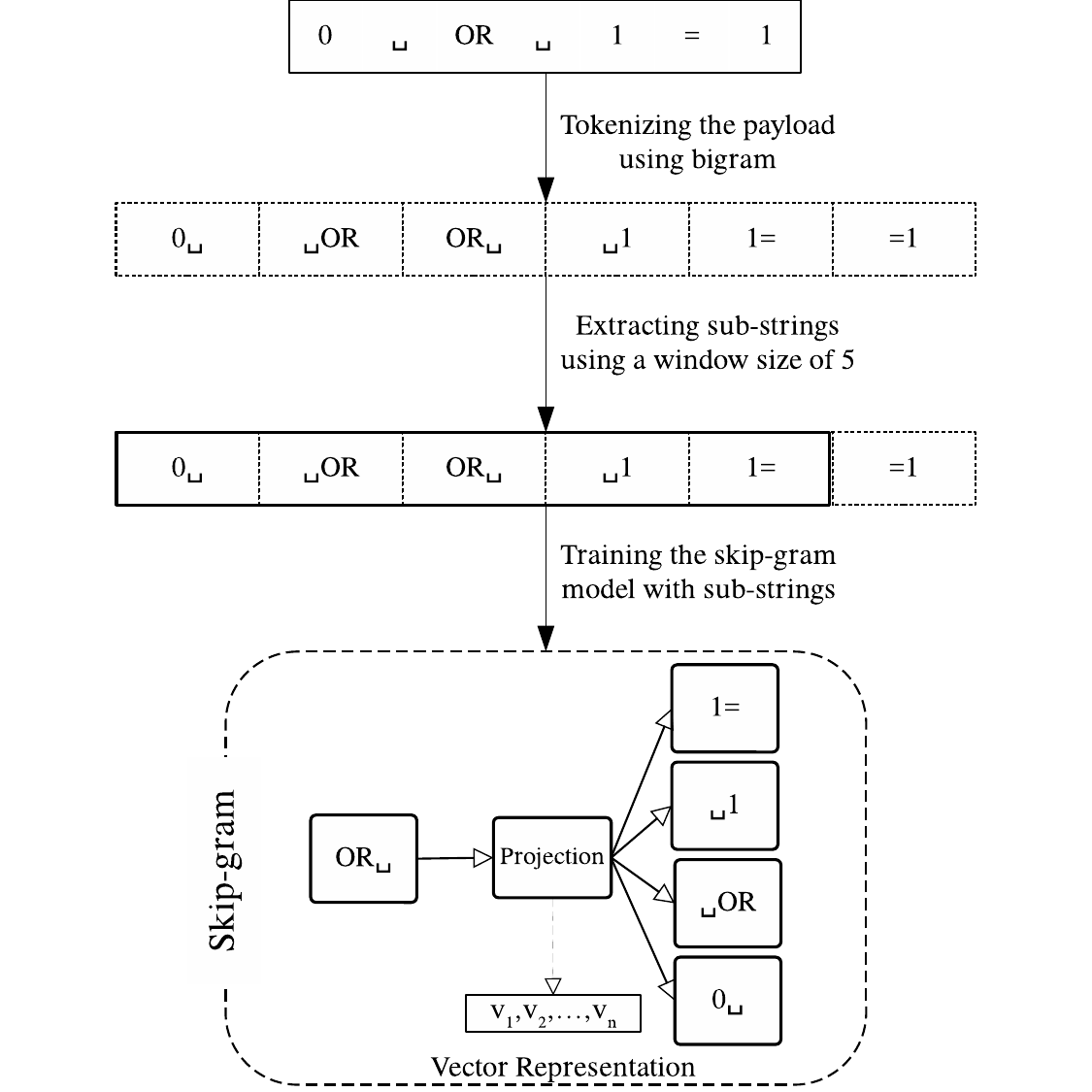}
  \caption{Example of learning numerical vector representation for $n$-grams of an SQLi attack payload using the skip-gram model.}
  \label{fig:we_example}
  \vspace{-2em}
\end{figure}
\subsubsection{Word Embedding} \label{we}
In code injection attacks different string fragments can be used interchangeably (i.e., in SQLi, \texttt{1=1} is an alternative for \texttt{"a"="a"}). In order to cluster similar attack payloads, we can consider alternative fragments as a single feature. For this purpose, we use word2vec \citep{mikolov2013efficient} to learn a vector representation for each string fragment; thus, later, we can measure the similarity between fragments. 

Continuous Bag of Words (CBOW) and skip-gram are two architectures of Word2Vec \citep{mikolov2013efficient}. These two simple neural network models are illustrated in \figurename~\ref{fig:we_models}. In both models, the middle layer is where the numerical vector is learnt for each unique word. CBOW learns this vector representation by using the surrounding words to predict the word in the middle. Opposed to CBOW, skip-gram tries to predict the surrounding words given the middle word. In our research, we observed that skip-gram could produce better word representation than CBOW. The problem with CBOW might be that since the target words are the output of the neural network, rare words have to compete with their alternatives that are repeated frequently. Therefore, rare words receive low attention from the model if the dataset is unbalanced. On the other hand, in skip-gram, the target words are the input of the neural network. Thus, rare words will not compete with frequent ones, and the model fairly learns the vector representation for every word. In this research, due to the rarity of fragments in our massive datasets, we use the skip-gram model.

In this research, each dataset is used to train a separate skip-gram model from scratch. \figurename~\ref{fig:we_example} depicts an example of transforming SQLi $n$-grams into numerical vectors. First, payloads are tokenized using $n$-gram. Then, a window with a specific size (see Section~\ref{param_embedding}) moves on the new strings to select sub-strings to train the skip-gram. Finally, skip-gram learns a vector representation for each $n$-gram token. In this specific example, the token \texttt{||} means \texttt{OR}, and the token \texttt{$\mathtt{\sim}$} means whitespace; thus, fragments \texttt{||$\mathtt{\sim}$} and \texttt{OR␣} are semantically the same, and they can be used interchangeably. Therefore, after completing the skip-gram training, these two tokens are expected to have similar vectors.

\subsubsection{Hierarchical Clustering} \label{hc}
After vectorizing the fragments, we use hierarchical clustering with cosine distance to cluster similar fragments.

Suppose vectors $v\textsubscript{1}$ and $v\textsubscript{2}$ represent fragments $f\textsubscript{1}$ and $f\textsubscript{2}$ respectively, the following equation is the distance calculation between $f\textsubscript{1}$ and $f\textsubscript{2}$.
\begin{equation} \label{eq:cosine}
\begin{split}
distance(f_1, f_2) & = 1 - \frac{v_1 . v_2}{\lVert v_1 \rVert_{l_2}\times\lVert v_2 \rVert_{l_2}}
\end{split}
\end{equation}

The output of \equationname~\eqref{eq:cosine} is a number in the range $[0,1]$. For two identical vectors, the calculated distance is $0$, and a calculated distance as $1$ indicates that two vectors are orthogonal.
Table~\ref{tab:binary} shows the binary vector for $p$.

\begin{table}[!t]
\caption{Example of a binary representation for a payload $p$, which shows whether the corresponding payload contains any fragment from the cluster $C_i$ where $i$ is the cluster's number.}
\label{tab:binary}
\begin{adjustbox}{center}
\renewcommand{\arraystretch}{1.5}
\begin{tabular}{|c|c|c|c|c|c|c|}
\hline
payload & $C_1$ & $C_2$ & $C_3$ & $C_4$ & $C_5$ & $C_6$ \\ \hline
$p$      & 1  & 0  & 1  & 1  & 0  & 0  \\ \hline
\end{tabular}
\end{adjustbox}
\vspace{-2em}
\end{table}

\subsubsection{Binary Encoder} \label{be}
Given a set of payload collection with clustered fragments, we transform each payload into a binary vector as an input to our clustering algorithm. Each item in our vector represents a cluster that indicates whether a corresponding payload contains any fragment belongs to that cluster. For instance, assume a payload $p$ contains a set of unique fragments $F=\{f_1, f_2, f_3, f_4, f_5\}$ and the set of obtained clusters from the previous step is $C=\{C_1,C_2,C_3,C_4,C_5,C_6\}$ in which $\{f_1,f_3\} \subseteq C_1$, $\{f_2\} \subseteq C_3$ and $\{f_4, f_5\} \subseteq C_4$. In Table~\ref{tab:binary}, the corresponding values for $C_2$, $C_5$, and $C_6$ are set as 0 indicates that $p$ does not contain any fragment belongs to these clusters.
\vspace{-0.5em}
\subsubsection{Deep Embedding Network (DEN)} \label{den}
As the last step of this section, we use the output of Binary Encoder to cluster attack payloads. The output of the Binary Encoder is a complex binary vector in which the items are respective to each other. Although numerous similarity measures have been proposed for binary features \citep{choi2010survey}, we found that clustering our high-dimensional raw data using these similarity measures results in poor performance. Thus, we considered using feature transformation methods to map our raw data to a much distinguishable feature space.

Mainly, data transformation methods include linear transformation such as Principal component analysis (PCA) \citep{wold1987principal} and non-linear transformation such as kernel methods \citep{hofmann2008kernel}. In recent years, the development of deep learning has facilitated the non-linear transformation of raw features into more clustering-friendly representation \citep{min2009deep}. Consequently, abundant deep learning-based clustering methods have been proposed in the literature \cite{chen2017unsupervised,yang2017towards,ghasedi2017deep,huang2014deep}.

In this research, we use DEN \citep{huang2014deep}, which first, utilizes a deep autoencoder to learn lower-dimensional representation from the input data. Then, it uses $k$-means to cluster learned features. In the following paragraphs, we detail the process.

\begin{figure}[!t]
\centering
  \includegraphics[width=\linewidth]{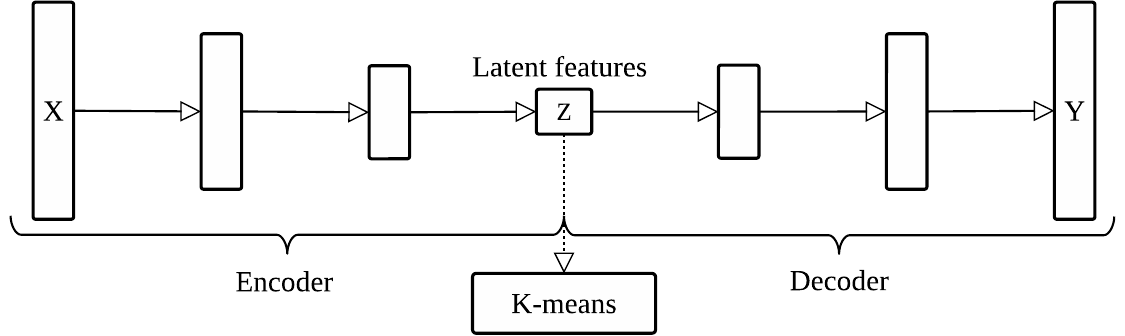}
  \caption{Architecture of Deep Embedding Network, we used to cluster our dataset. We feed the $k$-means algorithm with the output of the encoder.}
  \label{fig:den}
  \vspace{-2em}
\end{figure}

Autoencoder is a kind of unsupervised neural network consists of two parts: encoder function $h=f(\hat{x})$ and decoder function $r=g(h)$. The encoder maps raw data into a latent representation, and then the decoder reconstructs the raw input data from latent features (see \figurename~\ref{fig:den}). Simply, an autoencoder learns latent representation by minimizing reconstruction loss function $L(\hat{x}, g(f(\hat{x})))$. Considering that the input is binary, we use binary cross-entropy as the loss function. Given an input vector $\hat{x} = [x_1,x_2,\dots,x_n]$ and corresponding output vector $\hat{y} = [y_1,y_2,\dots,y_n]$, the loss calculation is
\begin{equation} \label{eq:cross-entropy}
\begin{split}
L(\hat{x}, g(f(\hat{x}))) & = - \frac{1}{n} \sum_{i=1}^{n} x_i \times \ln y_i + (1-x_i)\times\ln (1-y_i)
\end{split}
\end{equation}

In this paper, we use an autoencoder with six dense hidden layers to map features into three-dimensional feature space (see Section~\ref{ae} for more details). After training the autoencoder, we detach the decoder part and use the encoder to transform our data to a new clustering-friendly representation. Then, we cluster new features using the $k$-means algorithm for simplicity (see \figurename~\ref{fig:den}). It is to be noted that the impact of the clustering algorithm can be studied in further researches.

\vspace{-1em}
\subsection{Feature Extraction} \label{features}
At this stage, we pick only principal fragments from the fragment collection, which we have earlier created in Section~\ref{n-gram}. Then we calculate Inverse-Document-Frequency (IDF) for each fragment to build the final feature vector. We repeat this stage for each cluster individually.

\subsubsection{Token Selector} \label{ts}
Since each cluster is a subset of the original dataset, a cluster's fragment set is a subset of the main fragment collection. Therefore, as the first step, we remove unused fragments, and then we keep the remaining. For instance, let us assume that there are four fragments extracted from the whole dataset. However, payloads of a cluster are consisting of only two of them. Thus, since the adaptive searching in each cluster is done independently, we do not need those two extra fragments in that cluster. We, therefore, do not consider those unused fragments in building feature vectors in that cluster.

In order to improve our algorithm's performance, within each cluster, we remove non-informative fragments such as highly repetitive and exceedingly rare fragments from our fragment set. For this purpose, first, we calculate entropy for each fragment, and then we remove fragments with an entropy value lower than a threshold. The entropy calculation is given below
\begin{equation} \label{eq:entropy}
\begin{split}
E(f_i) & = -(p(f_i)\log_2 p(f_i) + (1-p(f_i)) \log_2 (1-p(f_i)) )\\  
&= -((\frac{k_i}{N}\times \log_2(\frac{k_i}{N})) + (\frac{N-k_i}{N}\times \log_2(\frac{N-k_i}{N})))
\end{split}
\end{equation}
where $i$ is the fragment's number, $E$ is the entropy, $p(f_i)$ is the probability of a fragment $f_i$ being present in a randomly selected payload, $k_i$ is the number of payloads containing the fragment $f_i$, and $N$ is the size of corresponding cluster.

\subsubsection{IDF Calculator} \label{idfcalculator}
Among the remaining fragments, rare ones are more valuable; thus, we calculate a weight for each unique fragment using inverse document frequency \citep{papineni2001inverse}. Then, we create a feature vector for each payload. If a payload contains a fragment, the value of the fragment in the feature vector will be its weight; otherwise, it will be zero. The IDF calculation is as follows:
\begin{equation} \label{eq:idf}
\begin{split}
w_f^i & = \ln(\frac{N}{k_i})
\end{split}
\end{equation}
where $i$ is the fragment's number, $w$ is the weight for fragment $f_i$, $N$ is the number of payloads, and $k_i$ is the number of payloads containing the fragment $f_i$ within the corresponding cluster.
\vspace{-1em}
\subsection{Test Oracle} \label{oracle}
Since our approach is a type of black-box testing, the only information we can obtain from a protector WAF is whether a request is identified as malicious or benign. Therefore, we propose an adaptive search technique; we call here AdaptiveSearch that benefits from prior experiences to minimize failed attempts. The key insight underlying AdaptiveSearch is that if a fragment attends more previously blocked attacks than others, it is more likely to cause a failure. Thus, if a new test payload contains these fragments, it is more likely to be recognized by the WAF. More specifically, in AdaptiveSearch, we assume that the set of previously blocked attacks is a document, and new test candidates are the queries. To find a payload with the highest likelihood of bypassing the WAF, for each query, we calculate a term frequency-inverse document frequency (TF-IDF) score that indicates the relevance of the query to the document \citep{jabri2018ranking}. Finally, we pick the payload with the lowest score and then execute it against the WAF. If the chosen payload bypasses the WAF, in further tests, we do not consider its fragments in our score calculation; otherwise, we add the payload to the document.

\begin{algorithm}[!t]
\caption{Adaptive Search algorithm}\label{searchalgorithm}
\begin{algorithmic}[1]
\Procedure{AdaptiveSearch}{$\textit{cluster}$, $\textit{rounds}$}
\State $\textit{P} \gets \textit{getPayloads(cluster)}$

\State $\textit{BV} \gets \textit{getBlockedVector(cluster)}$
\State $\textit{PV} \gets \textit{getBypassingVector(cluster)}$
\State $\textit{S} \gets \{\emptyset\} $
\State $\textit{SR} \gets \textit{+1}$\Comment{Reward}
\State $\textit{FR} \gets \textit{-0.5}$\Comment{Punishment}
\State $\textit{R} \gets \textit{0}$\Comment{Sum of Rewards}
\For {$\textit{i} = 1 \text{ to } \textit{rounds}$}

\If {$\textit{is first time}$}
\State $\textit{test\_candidate} \gets \textit{pickRandomPayload(P)}$
\Else
\State $\textit{rt} \gets \textit{RankPayloads(P, BV, PV)}$
\State $\textit{testCandidate} \gets \textit{pickBestPayload(rt)}$

\EndIf
\State $\textit{P} \gets \textit{P} - \textit{testCandidate}$
\State $\textit{result} \gets \textit{evaluate(testCandidate)}$
\If {$\text{\textit{result} is \textit{successful}}$}
\State $\textit{S} \gets \textit{S} \cup \{\textit{testCandidate}\} $
\State $\textit{PV} \gets \textit{updateBypassingVector(testCandidate)}$
\State $\textit{R} \gets \textit{R} + \textit{SR}$
\Else
\State $\textit{BV} \gets \textit{updateBlockedVector(testCandidate)}$
\State $\textit{R} \gets \textit{R} + \textit{FR}$
\EndIf
\EndFor
\State $\textit{saveClusterState(P, BV, PV)}$
\State \Return $\textit{S} , \textit{R}$
\EndProcedure
\end{algorithmic}
\end{algorithm}

The pseudo-code for AdaptiveSearch is given in Algorithm~\ref{searchalgorithm}. Line 2 initializes the payload collection of the cluster. Line 3 defines an array to keep the term frequency of the previously blocked attacks' fragments.  We also initialize a binary vector that keeps the state of each fragment (line 4). If a fragment attends a bypassing payload, its value will be zero; otherwise, it will be one. The value one for a fragment means that we count it in the score calculation.

At the very beginning of the search, we pick the first payload randomly (lines 10-12). In further searches, for each new attempt, we calculate the score of each payload, and then we pick the best payload (lines 13-14). The score calculation is given below
\begin{equation} \label{eq:tfidf}
\begin{split}
score(p) &= \sum_{i=1}^{n} b_i \times v_i
\end{split}
\end{equation}
where $i$ is the fragment's number, $b_i$ is the frequency of the fragment $f_i$ in the blocked attacks, $v$ is the feature vector for payload $p$, calculated in Section~\ref{idfcalculator}, and $n$ is the size of $v$.

Each time we pick a payload, we remove it from payload collection; then, we execute it against the WAF (lines 16-17). If a chosen attack bypasses the WAF, we add it to the bypassing collection (line 19). Then in the bypassing vector, we set the value of the items that represent a fragment belonging to the successful payload to zero (line 20). Otherwise, we update the failure vector (line 23) using the following equation:
\begin{equation} \label{eq:b_update}
\begin{split}
b_i &= s_i \times (tf_i + b_i)
\end{split}
\end{equation}
where $i$ is the fragment's number, $b_i$ is the frequency of the $i$th fragment in failure vector, $tf_i$ is its frequency in the payload $p$, and $s_i$ is the state value of fragment $f_i$ in success vector.

\begin{table}[!t]
\caption{Example of the payloads' term frequencies in cluster $c$.}
\label{tab:oracle_example}
\begin{adjustbox}{center}
\renewcommand{\arraystretch}{1.5}
\begin{tabular}{|p{0.03\textwidth}|p{0.02\textwidth}p{0.02\textwidth}p{0.02\textwidth}p{0.02\textwidth}|
p{0.04\textwidth}p{0.04\textwidth}p{0.04\textwidth}p{0.04\textwidth}|}
\hline
\multirow{2}{*}{p.id} & \multicolumn{4}{c|}{Term Frequencies} & \multicolumn{4}{c|}{Inverted Document Frequencies} \\ \cline{2-9} 
                          & f1        & f2       & f3       & f4       & f1           & f2           & f3           & f4          \\ \hline
1                         & 2         & 0        & 1        & 0        & 0.28         & 0            & 0.69         & 0           \\ \hline
2                         & 1         & 1        & 0        & 0        & 0.28         & 0.28         & 0            & 0           \\ \hline
3                         & 0         & 1        & 2        & 3        & 0            & 0.28         & 0.69         & 1.38        \\ \hline
4                         & 2         & 2        & 0        & 0        & 0.28         & 0.28         & 0            & 0           \\ \hline
\end{tabular}
\end{adjustbox}
\vspace{0em}
\end{table}
To put it in more perspective, let us assume that there are four attack payloads in cluster $c$. Table~\ref{tab:oracle_example} shows the frequencies of fragments in these payloads as well as their feature vectors. Assuming the first two payloads are tested and blocked, we sum their term frequency vectors to calculate the vector $\hat{b}$. As a result, the vector $\hat{b}$ is $\hat{b}=[3,1,1,0]$. To select the next payload to test, we calculate the dot product of b and the remaining payloads' feature vectors. The resulting ranks are: $Rank(p_3) = 0.97$ and $Rank(p_4) = 1.12$. Thus, we select the third payload, which has the lower rank. If $p_3$ bypasses the WAF, we set the values of fragments $f_2$, $f_3$ and $f_4$ in $\hat{s}$ vector to zero as $p_3$ contains these fragments. The resulting $\hat{s}$ is $\hat{s}=[1,0,0,0]$. Therefore, in the next update, frequencies of these fragments in $\hat{b}$ will be set to zero.

On lines 21 and 24, Algorithm~\ref{searchalgorithm} calculates a reward, which we explain in the following sections. Finally, it saves the current state of the cluster (line 27) and returns bypassing attacks and the reward (line 28).

In our observations, we realized that effective payloads are usually rare in payload collection; thus, only a few numbers of clusters contain bypassing payloads. Therefore, limiting searches to these clusters reduces the number of failed attempts significantly. For this purpose, we use the $\epsilon$-greedy policy for cluster selection.

\begin{algorithm}[!t]
\caption{Epsilon Greedy Policy}\label{EpsilonGreedy}
\begin{algorithmic}[1]
\Procedure{EpsilonGreedy}{}
\State $\textit{PC} \gets \textit{Payload Clusters}$
\State $\textit{S} \gets \{\emptyset\} $
\State $max\_\epsilon ,\epsilon \gets \textit{Epsilon}$
\State $\textit{R} \gets \textit{Number of Searching Rounds per episode}$
\State $\textit{K} \gets \textit{Update Rate for Epsilon}$
\State $\textit{E} \gets \textit{Episodes}$

\State $\textit{AR} \gets \textit{Initial Average Reward of Each Cluster}$

\For {$\textit{i} = 1 \text{ to } \textit{E}$}
\State $\textit{c} \gets \textit{pickCluster(}\epsilon\textit{)}$
\State $\textit{bypassing, reward} \gets \textit{ADAPTIVESEARCH(c, R)}$
\State $\textit{AR} \gets \textit{updateAverageReward(c, reward, AR)}$
\State $\textit{S} \gets \textit{S} \cup \textit{bypassing} $
\State $\epsilon \gets \textit{updateEpsilon(}max\_\epsilon\textit{, i, K)}$
\EndFor
\EndProcedure
\end{algorithmic}
\vspace{-0.25em}
\end{algorithm}
To use $\epsilon$-greedy policy in our method, we define the required terms and parameters as follow:
\begin{enumerate}
\item Reward: A positive constant value for each successful attempt.
\item Punishment: A negative constant value for each unsuccessful attempt.
\item Action: An action is the $R$ (line 5 of Algorithm \ref{EpsilonGreedy}) number of attempts that are made for a single cluster.
\item Action Reward: Sum of rewards and punishments per action.
\item Average Reward: Average of a cluster's action rewards.
\end{enumerate}

In $\epsilon$-greedy policy (Algorithm~\ref{EpsilonGreedy}), first, we value the actions based on their average rewards. Then we either select a random action with the probability of $\epsilon$ or the best action with the probability of $1 - \epsilon$ (line 10).

At the very beginning of the search, we do not have any information about clusters; thus, we set initial $\epsilon$ equal to a large number (e.g., $0.9$) to perform exploration. As we search more in clusters, we gain more knowledge. Therefore after each episode, we decrease the value of the $\epsilon$ to perform more exploitation (line 14). The calculation for $\epsilon$ reduction is given below:
\begin{equation} \label{eq:decay}
\begin{split}
\epsilon & = max\_\epsilon \times e^{-(k\times \tau)}
\end{split}
\end{equation}
where $max\_\epsilon$ is the maximum value of the $\epsilon$, $k$ is a constant value (see Section~\ref{epg}), and $\tau$ is the number of played episodes.
\vspace{-1em}
\section{Empirical Study} \label{empirical}
This section aims to evaluate our proposed method on its efficiency and effectiveness. In Section~\ref{questions}, we introduce the research questions. Section~\ref{procedure}, briefly explains the process of testing the WAFs. In Section~\ref{environment} we describe the experimental environment, including subject WAFs, datasets, and evaluation metrics. Section~\ref{parameters} demonstrates the algorithm parameters, and finally, Section~\ref{results} answers the research questions.
\vspace{-1em}
\subsection{Research questions} \label{questions}
In our empirical study, we aim to answer the following questions:
\begin{quote}
\emph{Q1: Does the choice of $n$-gram matter?}
\end{quote}
\begin{quote}
\emph{Q2: How does clustering affect the performance?}
\end{quote}
\begin{quote}
\emph{Q3: How does RAT compare with the state-of-the-art techniques?}
\end{quote}
\begin{quote}
\emph{Q4: Is the efficiency and effectiveness of RAT acceptable in practice?}
\end{quote}

\textit{Q1} and \textit{Q2} assess the effect of clustering and $n$-gram on the performance of our approach. \textit{Q3} compares our technique with state-of-the-art methods, and \textit{Q4} investigates whether the efficiency and effectiveness of our approach are acceptable in practice. The following sections answer these questions.
\vspace{-1em}
\subsection{Procedure} \label{procedure}
In our experiments, we only target the WAF itself, not the application behind it. Since WAFs are independent of the application under protection, for the HTTP key-value pairs, they validate values regardless of keys. Thus, we apply attacks through dummy keys as the HTTP query and the cookie. For instance, we consider the HTTP GET query key as \texttt{q}, and then we set an attack payload \lq\lq\texttt{0\%20or\%201=1\%23}\rq\rq as the query string for \texttt{q}. As a result, the URL for the sample request is \lq\lq\texttt{http://example.com/?q=0\%20or\%201=1\%23}\rq\rq. For each test, we either target the GET query parameter or the cookie. In our experiments for SQLi, we test both HTTP GET query parameter and cookie; and for XSS, we only test the GET parameter. It is worth mentioning that our method is independent of the communication protocol; thus, it can be implemented using different request types such as POST,  GET, and SOAP messages. In this research, since the subject WAFs' rules are the same for both GET and POST requests, we conduct our experiments using GET for simplicity.

It is to be noted that when WAF receives a request, it investigates the request and if it detects an attack, responds that the request is forbidden. Thus, we understand that the attempt is failed; otherwise, we mark the attack pattern as bypassing.
\vspace{-0.5em}
\subsection{Experimental environment} \label{environment}
\textit{RAT} consist of two parts: Data Processor and Test Oracle. The Data Processor is a one time process and requires at least 32GB of RAM and a CUDA-enabled Graphic Card with the minimum compute capability 3.0 whereas Test Oracle requires the minimum 8GB of RAM with 2.10GHz duo core CPU. However, to speed up tests, all the experiments were conducted on a Server with two 2.10GHz Intel(R) Xenon(R) processors and 64GB RAM running Windows 10 pro. The program codes were written in Python 3.6, and the source code is publicly available on GitHub\footnote{\url{https://github.com/mhamouei/rat}}. Furthermore, the deep autoencoder implemented with Keras-GPU 2.2.4 running on the top of Tensorflow 1.9 and was executed on Google Colab\footnote{\url{https://colab.research.google.com}}. The following sections provide details about Subject WAFs, Datasets and metrics.
\subsubsection{Subject WAFs}
In our case studies, we apply our tests on a custom-built WAF and two famous open-source WAFs: \textit{ModSecurity}\footnote{\url{https://modsecurity.org}} and \textit{Naxsi}.\footnote{\url{https://www.nbs-system.com}}

\textit{ModSecurity} is a toolkit that provides real-time protection for web applications. It protects web applications against various types of attacks, such as SQLi, XSS, and denial of service. We deployed the \textit{ModSecurity} with an Apache HTTP server on a local virtual machine.

\textit{Naxsi} stands for \lq\lq Nginx Anti XSS and SQL Injection,\rq\rq which is a third-party module for Nginx web server that protects web applications against SQLi and XSS attacks. Similar to  \textit{ModSecurity}, we deployed \textit{Naxsi} on a local virtual machine.

Custom-built WAF is the modified version of ModSecurity, customized to protect a real-world application's web services. This application provides various educational services, and the private data of thousands of students is stored in its database. Therefore, custom-built WAF is responsible for the privacy of students' data.

\begin{table}[!t]

\caption{Number of samples in each dataset.}
\label{tab:dataset}

\begin{adjustbox}{center}
\renewcommand{\arraystretch}{1.5}
\begin{tabular}{p{0.3\textwidth}
p{0.3\textwidth}
p{0.3\textwidth}}
\hline
\rowcolor[HTML]{EFEFEF} 
Dataset name         & \multicolumn{2}{c}{\cellcolor[HTML]{EFEFEF}Number of Payloads} \\ \hline
SQL Injection        & \multicolumn{2}{c}{2,417,720}                    \\ \hline
Cross-site Scripting & \multicolumn{2}{c}{1,798,062}                    \\ \hline
\end{tabular}
\end{adjustbox}
\vspace{-0.0em}
\end{table}

\subsubsection{Datasets} \label{sec-dataset}
In this research, we evaluate our technique on two different injection datasets (available on GitHub\footnote{\url{https://github.com/mhamouei/rat_datasets}}). The first dataset is a collection of SQLi payloads which we generated using the finite BNF grammar proposed in \citep{appelt2018machine}; thus, we can fairly compare \textit{RAT} with \textit{ML-Driven E}. The next dataset is a collection of XSS payloads, we generated using an opensource fuzzer tool, named dharma.\footnote{\url{https://github.com/MozillaSecurity/dharma}} Table~\ref{tab:dataset} shows the number of payloads in each dataset.

\begin{table}[!t]
\caption{Percentage of bypassing payloads for each open-source WAF.}
\label{tab:bypassing}
\begin{adjustbox}{center}
\renewcommand{\arraystretch}{1.5}
\begin{tabular}{|c|c|c|c|}
\hline
\multirow{2}{*}{WAF} & \multicolumn{2}{c|}{SQLi} & XSS           \\ \cline{2-4} 
                     & GET Parameter   & Cookie  & GET Parameter \\ \hline
ModSecurity          & 0.007\%             & 0.079\%     & 0.015\%             \\ \hline
NAXSI                & 0.005\%             & 0.005\%     & 0.004\%             \\ \hline
\end{tabular}
\end{adjustbox}
\vspace{-2em}
\end{table}

\subsubsection{Effectiveness metrics}
This research aims to increase the number of discovered bypassing attacks while reducing failed attempts. Thus, we consider bypassing payloads as positives and blocked payloads as negatives. As a result of this naming, successful attempts are True Positives (TPs), and unsuccessful ones are False Positives (FPs).

To evaluate and compare the effectiveness of \textit{RAT} and \textit{Ml-Driven E}, we measure TP over the limited number of requests. Since deploying \textit{NAXI} and \textit{ModSecurity} on the local machines provided the feasibility of brute-force attacks, we applied brute-force attacks on these open-source WAFs by exhaustively testing all attack payloads of our collections on the WAFs. Then, we measured the number of positives for each request parameter. Table~\ref{tab:bypassing} shows the percentage of bypassing attack payloads existing in our collections for both WAFs. Knowing the total number of positives, we can calculate $\frac{TP}{positives}$ as True Positive Rate (TPR) or open-source WAFs. However, since communication with the custom-built WAF is through the Internet, we cannot calculate positives for the custom-built WAF. Thus, in our experiments, we only report the number of TPs for the custom-built WAF. Furthermore, to compare \textit{RAT} with \textit{ART4SQLi} and \textit{XSSART}, we need to measure the number of FPs before finding the first TP. The lower FP shows that the approach is faster in discovering the first bypassing payload.

\subsubsection{Efficiency metrics}
To evaluate the efficiency of \textit{RAT}, we measure the time spent per request (TSR). For this purpose, in each episode, we record the time used in each cluster, then we divide it by the total number of requests made from that cluster. Finally, we report the average TSR.
\vspace{-1em}
\subsection{Parameter settings} \label{parameters}
For each step, there are various parameters to set. For the $n$-gram tokenizer, the only parameter to set is the size of $n$, which we discuss in detail in Section~\ref{n_gram_setting}. The binary encoder does not have any parameters, and the remaining parameters are as follows:

\begin{table*}[!t]
\caption{Architecture of the AutoEncoder.}
\label{tab:auto_encoder}
\begin{adjustbox}{center}
\renewcommand{\arraystretch}{1.5}
\begin{tabular}{|c|c|c|c|c|c|c|c|c|c|}
\hline
\multirow{2}{*}{Attack Type} &
  \multirow{2}{*}{Input layer} &
  \multicolumn{3}{c|}{Encoder} &
  \multirow{2}{*}{Bottleneck} &
  \multicolumn{3}{c|}{Decoder} &
  \multirow{2}{*}{Output layer} \\ \cline{3-5} \cline{7-9}
     &    & \multicolumn{3}{c|}{Hidden layers} &   & \multicolumn{3}{c|}{Hidden layers} &    \\ \hline
SQLi & 52 & 36        & 23       & 10       & 3 & 10        & 23       & 36       & 52 \\ \hline
XSS  & 347 & 261        & 175       & 89       & 3 & 89        & 175       & 261       & 347 \\ \hline
\end{tabular}
\end{adjustbox}
\vspace{-1.5em}
\end{table*}
\vspace{-0.5em}
\subsubsection{Hierarchical Clustering} \label{param_hierarchical_clustering}
In this step, we need to set a threshold to cluster tokens with a dissimilarity of less than the threshold. Clustering more tokens together results in losing details, and clustering fewer tokens keeps more details, directly affect payload clusters quality. In this step, we do not want to either lose or keep too many details. In our experiments, we observed that a threshold between 0.2 to 0.5 could be a good choice. However, we obtained the best performance in discovering bypassing payloads by setting this threshold to 0.3, which means that if the cosine similarity between the tokens is more than 70\%, they should cluster together.
\vspace{-0.5em}
\subsubsection{Skip-gram} \label{param_embedding}
For the skip-gram, we used a Python library named Gensim\footnote{\url{https://radimrehurek.com/gensim}} with the almost default parameter settings. We only set the window size considering the value of $n$ in $n$-gram. For example, for $n=2$, we set the window size to 5. It is because when $n=2$, half of the middle token is repeated in the prior token and the other half is repeated in the posterior token. Thus, increasing the window size by two results in more meaningful training of the skip-gram as there is no intersection between these two tokens and the middle token (see \figurename~\ref{fig:we_example}).
\vspace{-0.5em}
\subsubsection{AutoEncoder Architecture} \label{ae}
In this research, we use the simplest possible AutoEncoder architecture to extract features for our purpose. Since the AutoEncoders' inputs are low-dimensional binary vectors, a deep AutoEncoder with dense layers can achieve high reconstruction accuracy. However, the choice of hyperparameters of deep learning models, such as the number of hidden layers and their size, significantly affect accuracy. Since the way of tuning these parameters is still an open problem \citep{shaziya2021impact}, we tuned AutoEncoder manually. In the tuning process, we observed that three hidden layers in each encoder and decoder parts are enough, and we could achieve slightly better accuracy by reducing and increasing the size of hidden layers uniformly. We tend to produce feature vectors with the lowest possible dimensionality for easier and better clustering. In our experiments, we attained a reasonable reconstruction accuracy (95.92\% and 96.76\% for XSS and SQLi datasets, respectively) by reducing the dimensionality to 3. The final architectures of the AutoEncoders are shown in Table~\ref{tab:auto_encoder}.
\vspace{-0.5em}
\begin{figure}[t!]

\subfloat[SQLi]{%
  \label{fig:silhouette_sqli}\includegraphics[width=0.5\columnwidth]{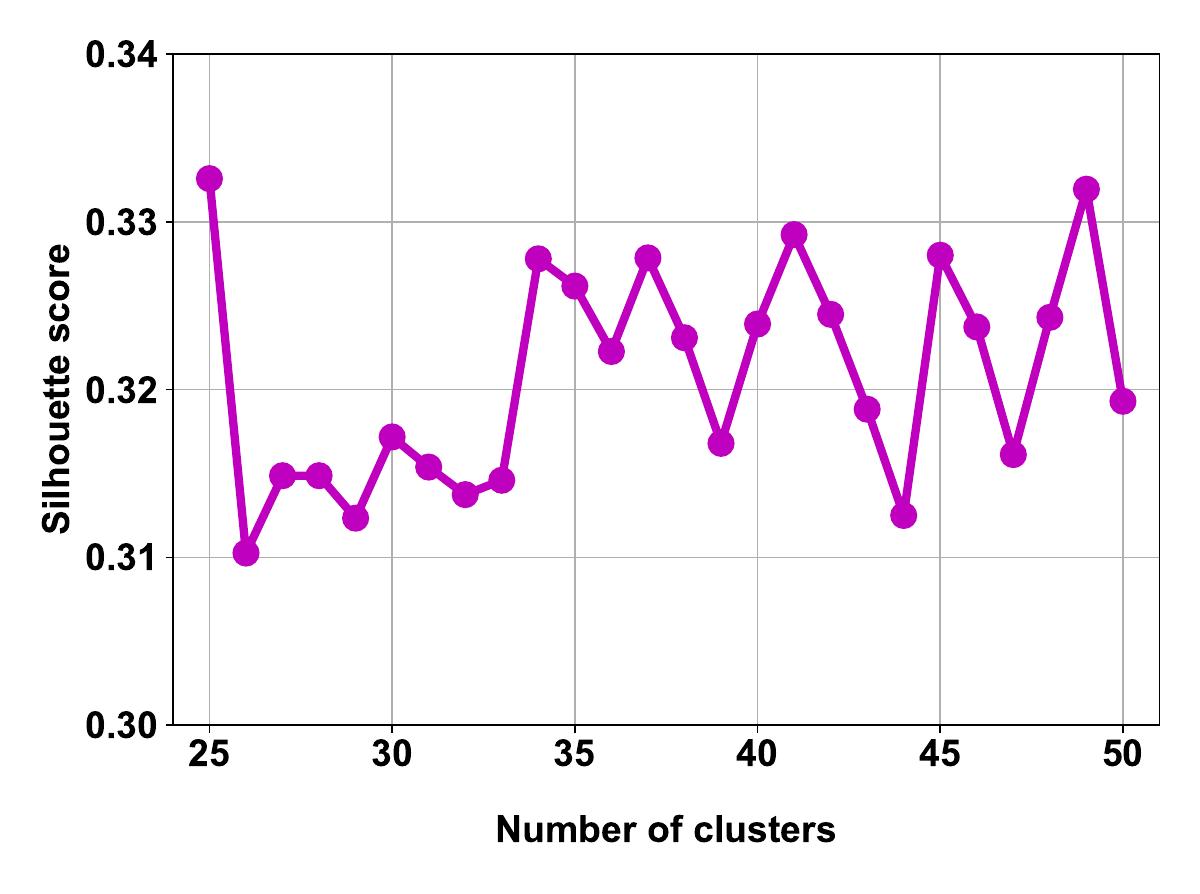}%
}
\hfill
\subfloat[XSS]{%
  \label{fig:silhouette_xss}\includegraphics[width=0.5\columnwidth]{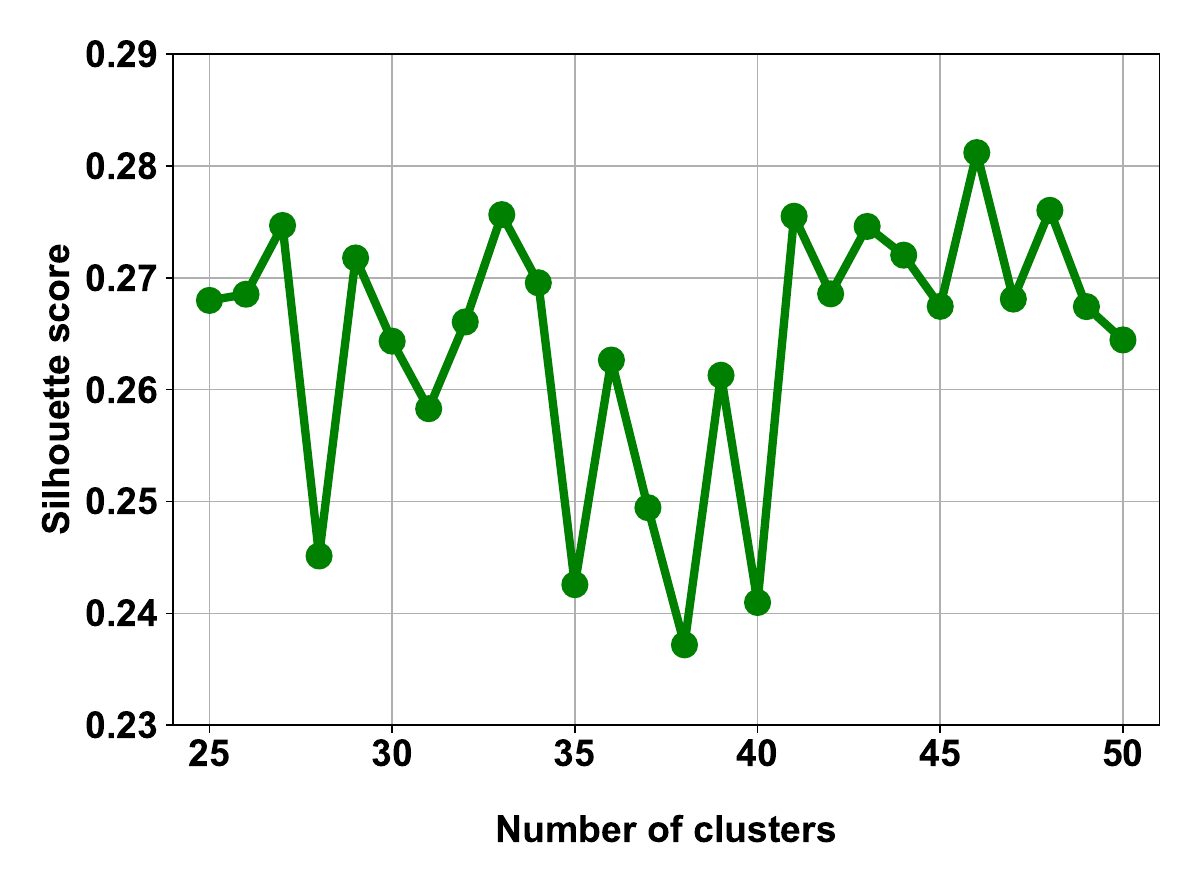}%
}
\caption{Silhouette scores for different $k$ in $k$-means in clustering SQLi and XSS datasets.}
\label{fig:silhouette}
\vspace{-2em}
\end{figure}
\subsubsection{Number of Clusters}
The time complexity for selecting each test candidate is $O(n)\times O_s$, where $n$ is the size of the cluster and $O_s$ is the time complexity of the \textit{Rank} function. Therefore, searching inside small clusters requires fewer computations than large clusters. Increasing the number of clusters reduces cluster size and, consequently, features; thus, our adaptive search algorithm would also require fewer searches inside smaller clusters to find bypassing payloads. However, having too many clusters poses two major problems:
\begin{enumerate}
\item Bypassing samples spread over more clusters, resulting in the rarity of bypassing payloads inside clusters. Thus, our search technique effectiveness drops.
\item $\epsilon$-greedy algorithm requires more exploration to filter clusters, and due to the previous problem, it is harder to escape local optima.
\end{enumerate}
In our observations, we realized that a number between 25 to 50 clusters is a good compromise for our datasets' size. To find the exact number of clusters, we used silhouette score \citep{rousseeuw1987silhouettes}, representing how good clusters are apart and distinguished from each other. The score range is between -1 to 1, and the higher score means that the clusters are better apart and distinguished. We clustered both datasets with all numbers between 25 and 50. We then measured the silhouette score for these clusters and picked the best clusters. \figurename~\ref{fig:silhouette} shows the Silhouette scores for the different number of clusters in clustering SQLi and XSS. In \figurename~\ref{fig:silhouette_sqli}, 25 clusters achieved the highest score, and in \figurename~\ref{fig:silhouette_xss}, the highest score belongs to 46 clusters. Therefore, we divide the SQLi and XSS datasets into 25 and 46 clusters, respectively.

\subsubsection{Entropy threshold}
For the final step of feature reduction, we use entropy to remove improper features. Note that an inadequate threshold value causes the improper features to remain, and a high threshold value results in information loss. In our experiments, with trial-and-error, we found the threshold of $T=0.05$ to be the optimal value.
\vspace{-1em}
\subsubsection{$\epsilon$-greedy parameters} \label{epg}
For the $\epsilon$-greedy algorithm, we set the reward to 1, and since we expect to experience failure more than success, we set the punishment to half of the reward value. We set the maximum $\epsilon$ to 0.9, and for the epsilon reduction (\equationname~\eqref{eq:decay}), we set the $k$ to \num{5e-3}. The value of $k$  controls the exploration and exploitation rate by adjusting the epsilon reduction speed. In other words, a high $k$ means RAT spends more episodes exploring clusters than a low $k$. We calibrated $k$ with trial-and-error.

\vspace{-1.5em}
\section{Results} \label{results}
In this section, we answer the research questions described in Section~\ref{questions} by our experiments.

\begin{figure}[t!]

\subfloat[ModSecurity]{%
  \includegraphics[clip,width=0.5\columnwidth]{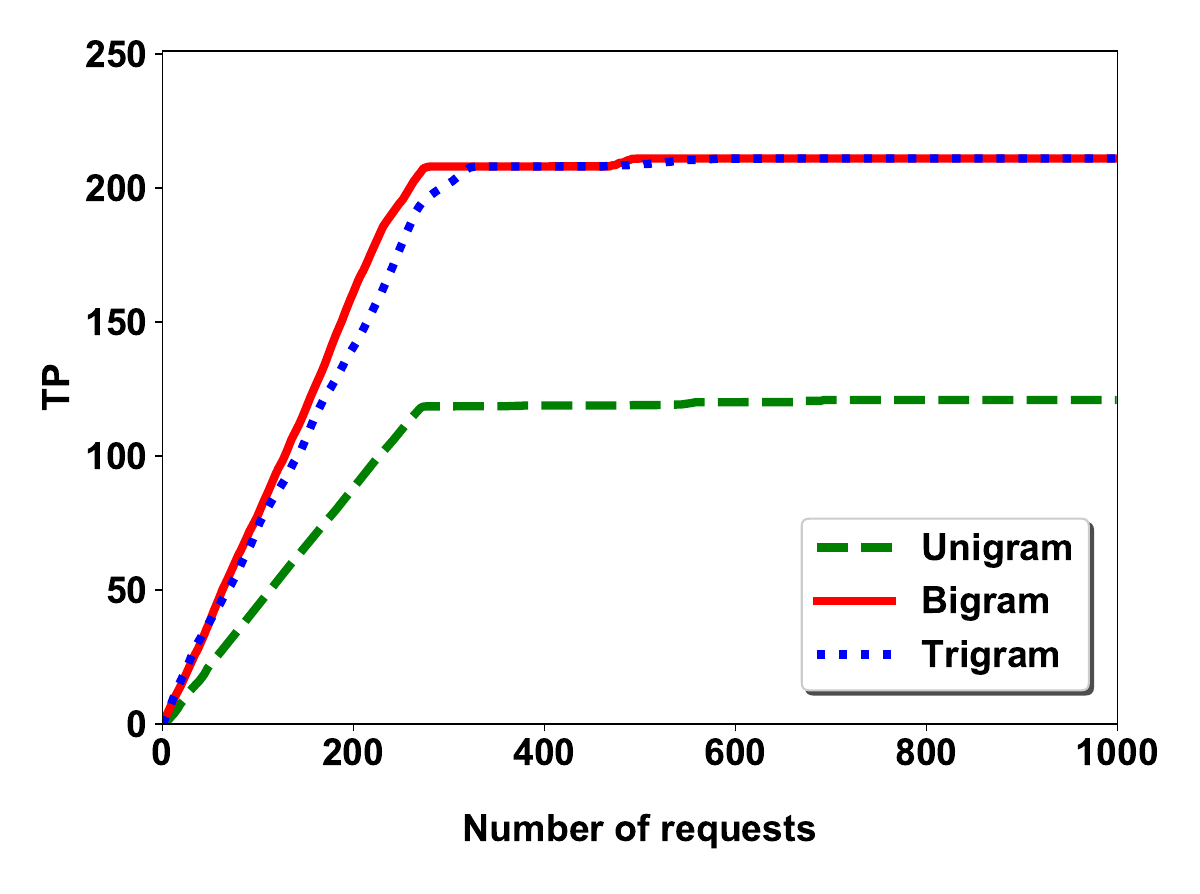}%
}
\hfil
\subfloat[NAXSI]{%
  \label{sqli_sub2}\includegraphics[clip,width=0.5\columnwidth]{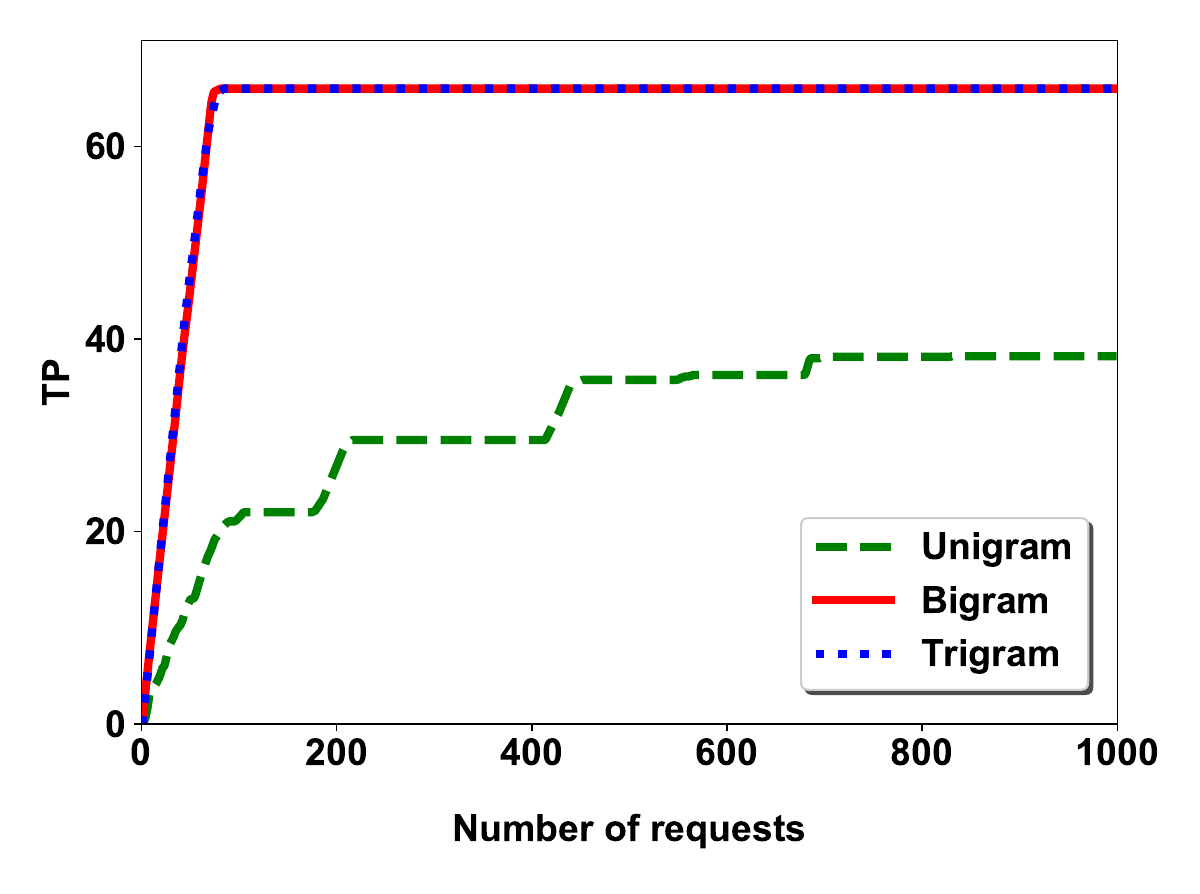}%
}
\centering
\subfloat[Custom-built WAF]{%
  \label{sqli_sub3} \includegraphics[clip,width=0.5\columnwidth]{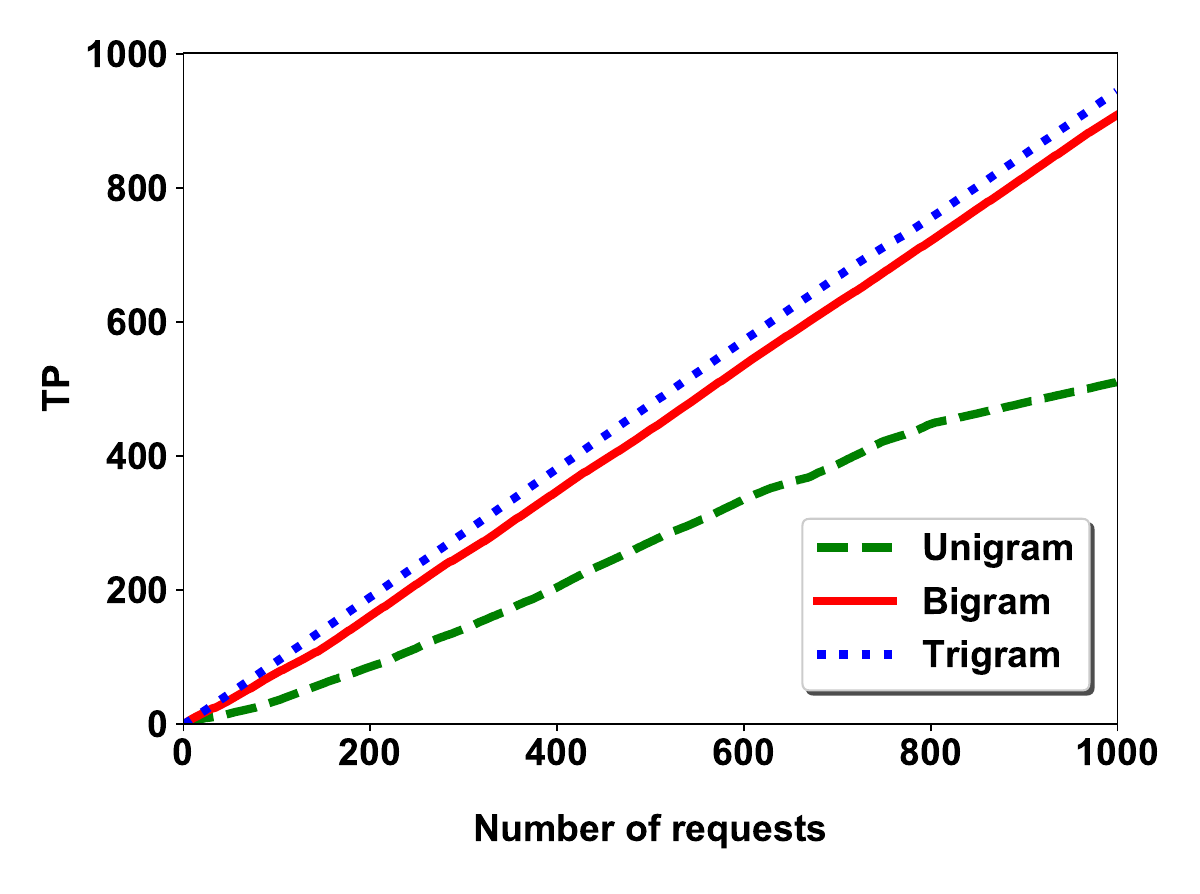}%
}
\caption{Average true positive in testing different WAFs for SQLi vulnerabilities with different values for $n$.}
\label{fig:sqli_ngram}
\vspace{-1em}
\end{figure}

\begin{figure}[t!]

\subfloat[ModSecurity]{%
  \includegraphics[clip,width=0.5\columnwidth]{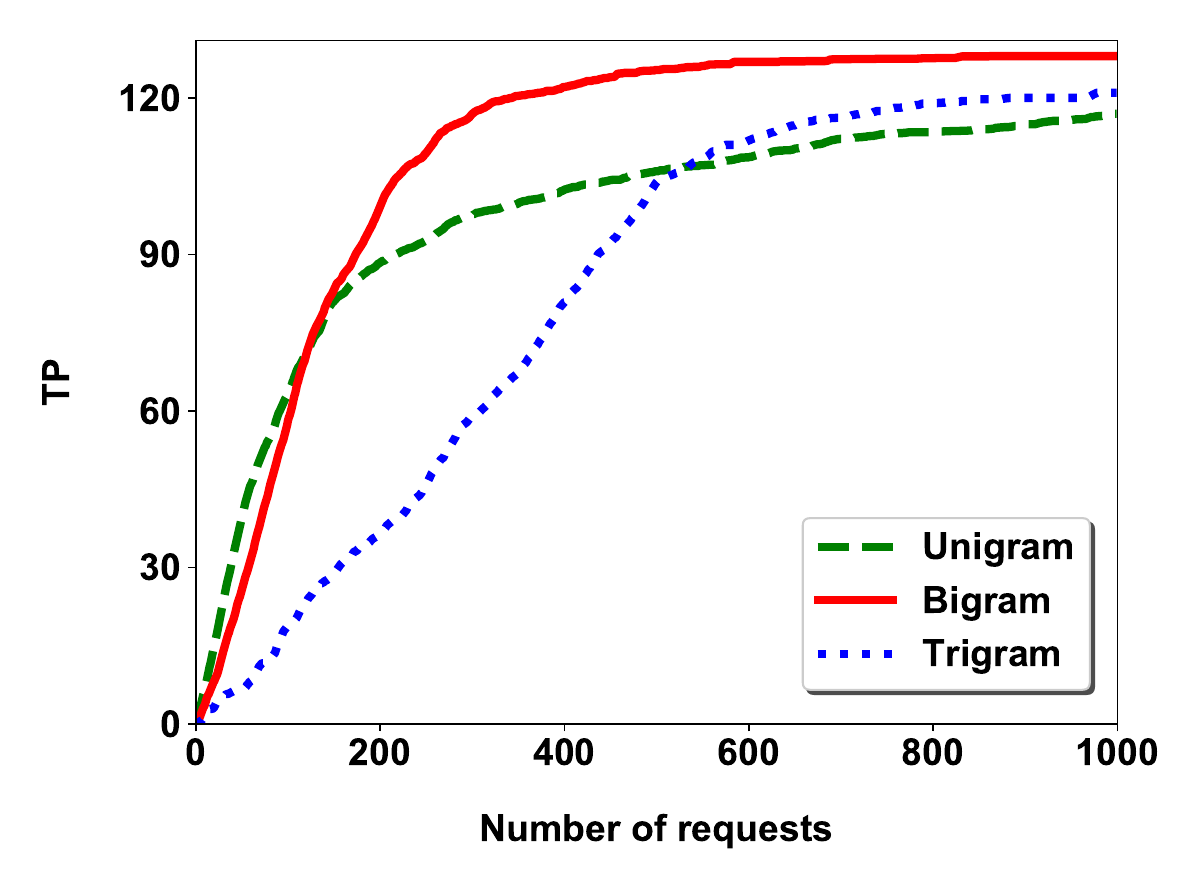}%
}
\hfil
\subfloat[NAXSI]{%
  \includegraphics[clip,width=0.5\columnwidth]{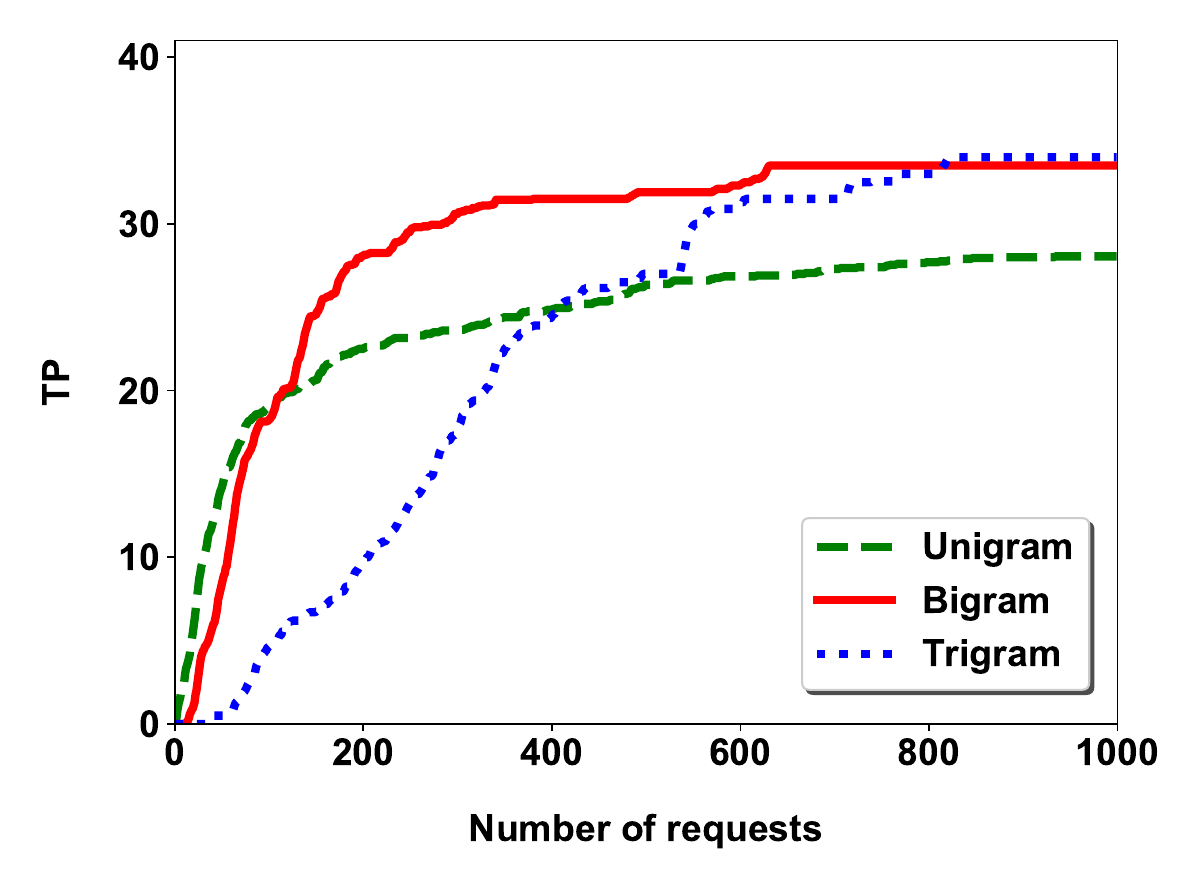}%
}
\centering
\subfloat[Custom-built WAF]{%
  \includegraphics[clip,width=0.5\columnwidth]{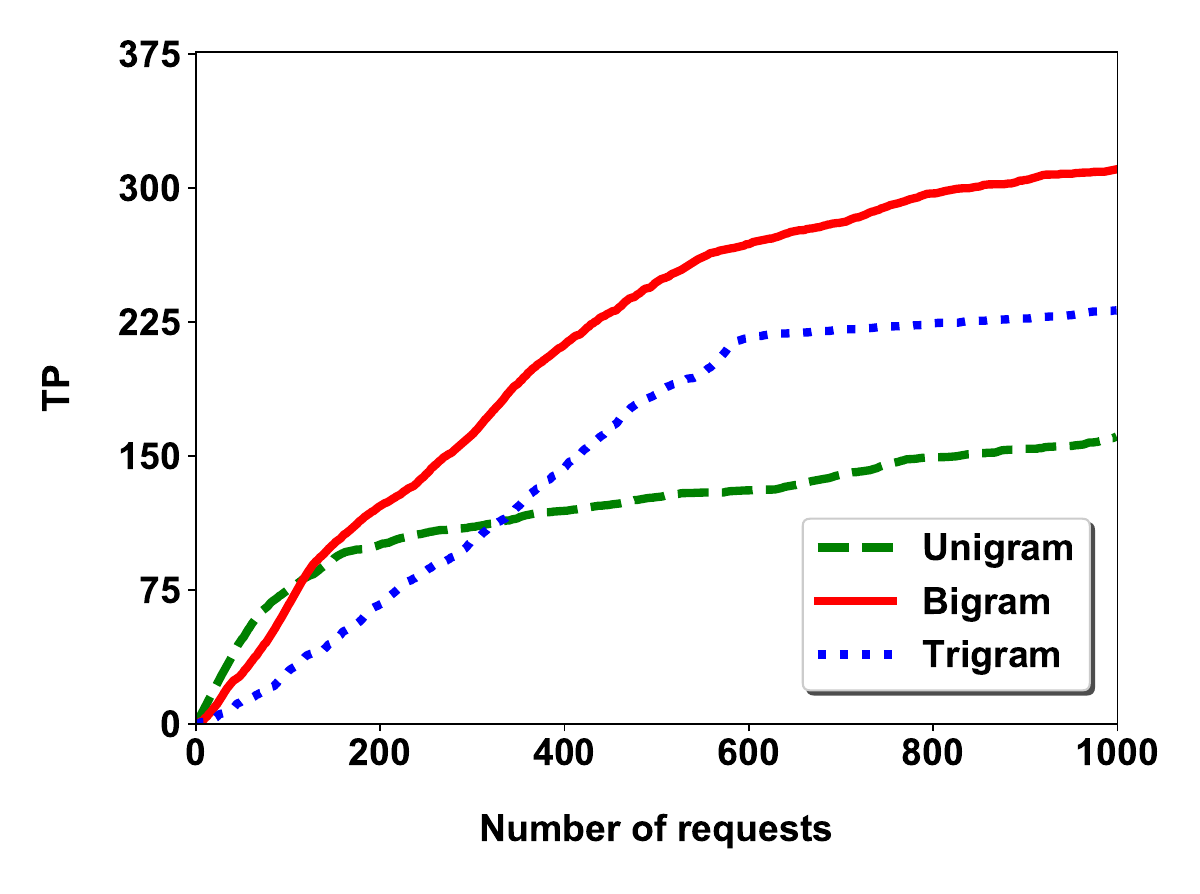}%
}
\caption{Average true positive in testing different WAFs for XSS vulnerabilities with different values for $n$.}
\label{fig:xss_ngram}
\vspace{-1em}
\end{figure}

\vspace{-1em}
\subsection{\textit{Q1: Does the choice of $n$-gram matter?}} \label{n_gram_setting}
To answer Q1, first, we picked four random subsets with the size of 25000 payloads containing bypassing payloads from both datasets (two subsets for each dataset). We then performed AdaptiveSearch on them using unigram, bigram, and trigram (100 repetitions for each). Then, we calculated the average TP over 1000 requests.

\figurename~\ref{fig:sqli_ngram} shows the results of applying AdaptiveSearch on the SQLi subsets, and \figurename~\ref{fig:xss_ngram} shows the results of the same experiment on the XSS subsets. Obtained results show that overall, unigram has poor performance as it can not models sophisticated patterns. We also observed that on the whole, the bigram was slightly more efficient than trigram. This is because although trigram extracts more sophisticated patterns than bigram, the number of distinct patterns extracted by bigram is fewer than trigram. Thus, bigram requires fewer observations than trigram. Moreover, the results show that bigram offers more robust results than others when testing different WAFs using different datasets. Therefore, in our further experiments, we used bigram for both SQLi and XSS datasets.

\begin{figure*}[t!]

\subfloat[SQLi]{%
  \includegraphics[width=\columnwidth]{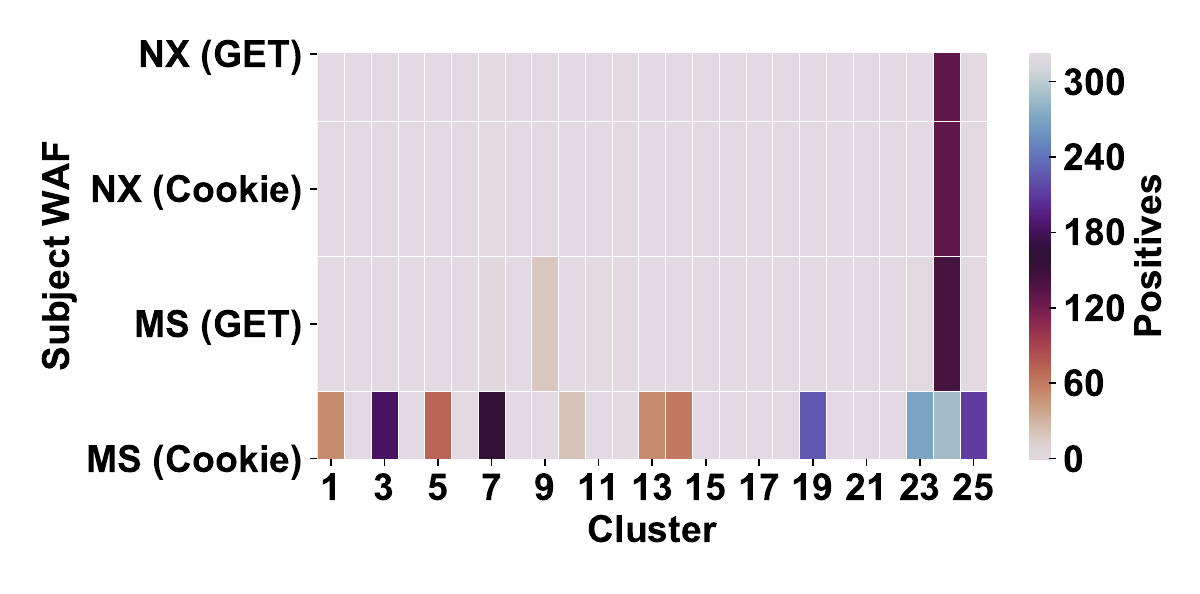}%
}
\hfill
\subfloat[XSS\label{fig:xss_dist}]{%
  \includegraphics[width=\columnwidth]{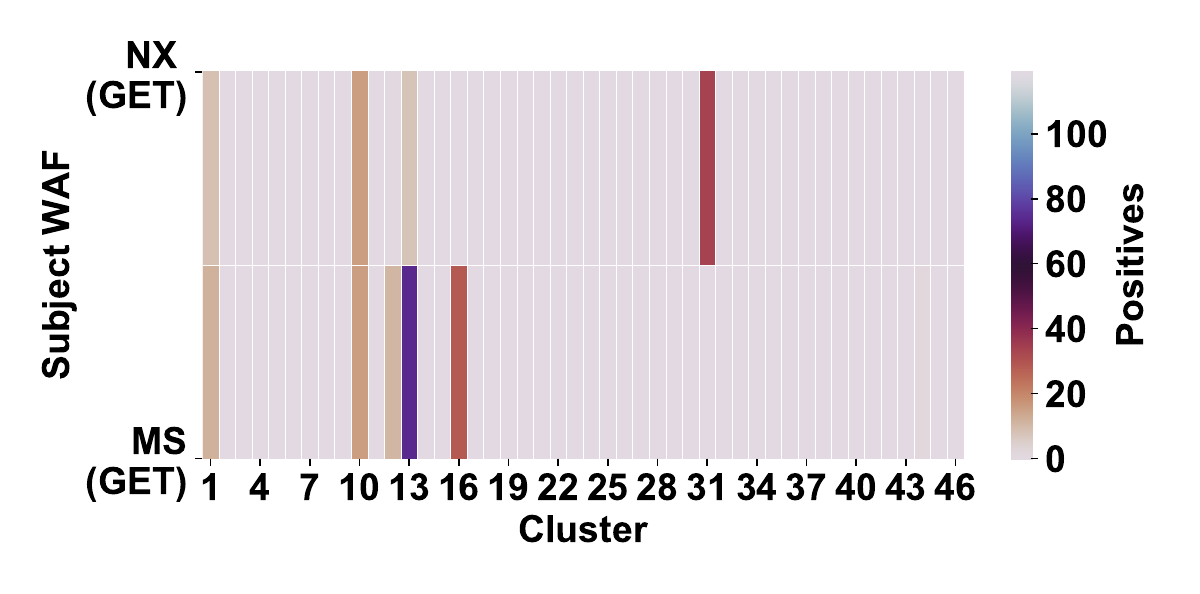}%
}

\caption{Distribution of bypassing payloads (TP) over clusters.}
\label{fig:bypassing_distro}
\vspace{-0.0em}
\end{figure*}
\begin{table*}[!t]
\caption{Number of test parameters before and after clustering.}
\label{tab:c_effect}
\begin{adjustbox}{center}
\renewcommand{\arraystretch}{1.5}
\begin{tabular}{|c|c|c|c|c|c|c|c|c|}
\hline
\multirow{3}{*}{Attack Type} & Before Clustering                 & \multicolumn{7}{c|}{After Clustering}                                               \\ \cline{2-9} 
                             & \multirow{2}{*}{Total Parameters} & \multicolumn{3}{c|}{$T = 0$} & \multicolumn{3}{c|}{$T = 0.05$} & \multirow{2}{*}{Ratio} \\ \cline{3-8}
     &        & min  & max    & mean   & min & max  & mean &     \\ \hline
SQLi & 185    & 51   & 173    & 73     & 49  & 126  & 69   & 247 \\ \hline
XSS  & 113408 & 826 & 48470 & 20714 & 103 & 1702 & 933 & 3.13 \\ \hline
\end{tabular}
\end{adjustbox}
\vspace{-2em}
\end{table*}
\vspace{-1em}
\subsection{\textit{Q2: How does clustering affect the performance?}}
To answer Q2, we investigated the effects of clustering from two aspects. The first aspect is that decreasing the number of test parameters reduces the number of observations. Thus we measured the number of fragments before and after the clustering phase. As shown in Table~\ref{tab:c_effect}, in the worst case, before applying the feature reduction ($t=0$), the clustering phase reduced the number of test parameters by 6.48\% and 57.26\% for SQLi and XSS datasets, respectively. With the clustering and feature reduction with the entropy threshold of $t=0.05$, in the worst case, we could reduce the number of parameters by 31.89\% and 98.50\% for SQLi and XSS datasets, respectively. We also measured the ratio $\frac{\text{number of samples}}{\text{number of test parameters}}$ for each cluster to check whether the ratio between samples and test parameters is acceptable. The worst ratio for each dataset is reported in the last column of Table~\ref{tab:c_effect}.

The second aspect is that bypassing attacks tend to cluster together; thus, only a limited number of clusters contain bypassing samples. To prove our claim, we applied brute force attacks on both Naxsi (NX) and Modsecurity (MS). Targeted parameters for the SQLi attack are a random HTTP GET parameter and Cookie, and for the XSS attack, we targeted a random HTTP GET parameter. Then, we analyzed the distribution of bypassing attacks within the clusters for each tested parameter.

The obtained results are illustrated in \figurename~\ref{fig:bypassing_distro}. In this figure, each column represents the number of a cluster's bypassing payloads in testing the corresponding parameter and WAF. For instance, in \figurename~\ref{fig:xss_dist}, the \nth{31} cluster contains more than 100 bypassing payloads in testing the HTTP GET parameter protected by \textit{ModSecurity}. The same cluster has about 40 bypassing payloads testing the same parameter protected by \textit{NAXSI}.

According to \figurename~\ref{fig:bypassing_distro}, bypassing samples are distributed within a few clusters. Therefore, finding effective clusters using $\epsilon$-greedy policy can significantly reduce unsuccessful attempts in uncovering bypassing payloads by discarding ineffective clusters.
\vspace{-1em}
\subsection{\textit{Q3: How does RAT compare with the state-of-the-art techniques?}}

To answer Q3, first, we implemented \textit{Ml-Driven E}\footnote{\url{https://github.com/mhamouei/ml-driven}} \citep{appelt2018machine}, \textit{ART4SQLi}\footnote{\url{https://github.com/mhamouei/art4sqli}} \citep{zhang2019art4sqli} and \textit{XSSART}\footnote{\url{https://github.com/mhamouei/xssart}} \citep{lv2019adaptive} based on the original papers, and then we compared \textit{RAT} with \textit{Ml-Driven E}, \textit{ART4SQLi} and a \textit{Random Fuzzer} using the SQLi dataset, and \textit{XSSART} using the XSS dataset. The comparison tests are designed concerning the objective of each approach. Moreover, to verify that \textit{RAT} also works with other attacks, we compare \textit{RAT} with a \textit{Random Fuzzer} using the XSS dataset.
\begin{figure}[t!]

\subfloat[ModSecurity]{%
  \includegraphics[width=0.5\columnwidth]{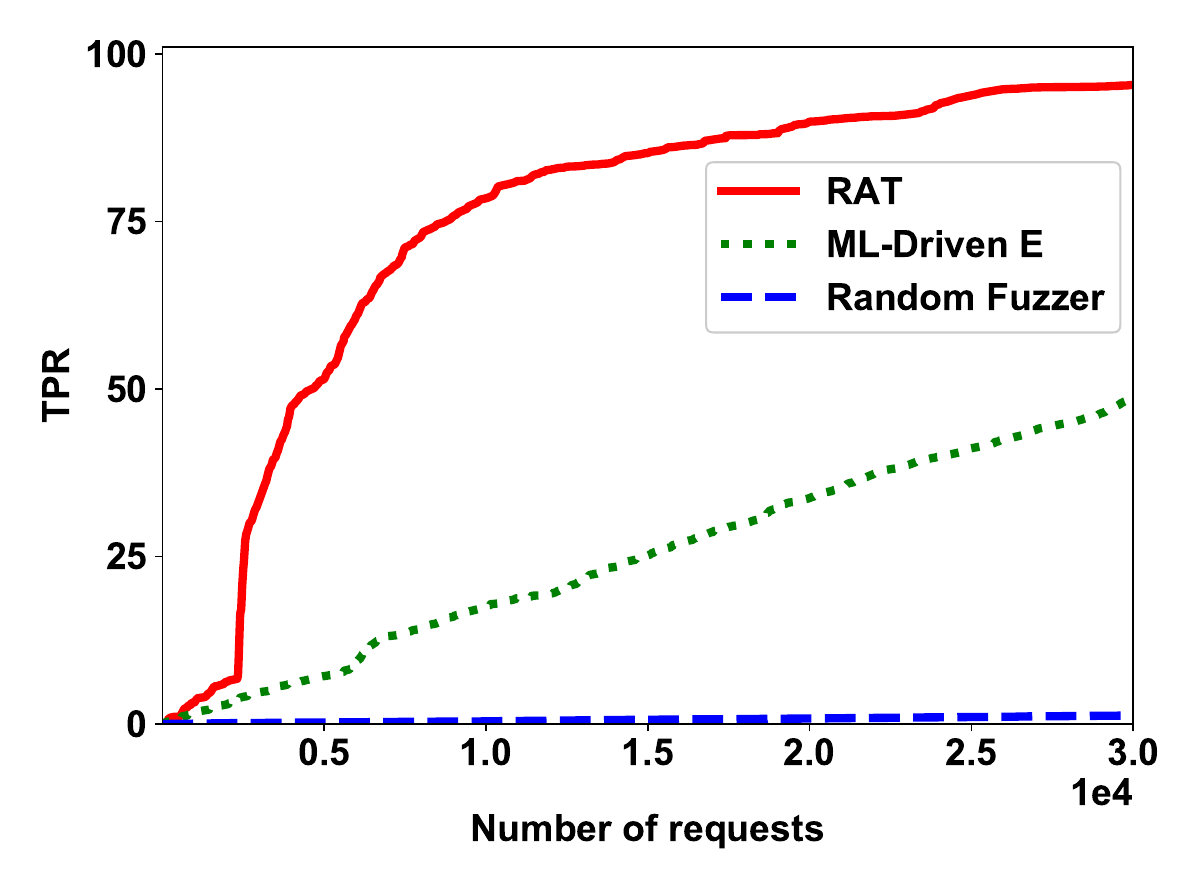}%
}
\hfill
\subfloat[NAXSI]{%
  \includegraphics[width=0.5\columnwidth]{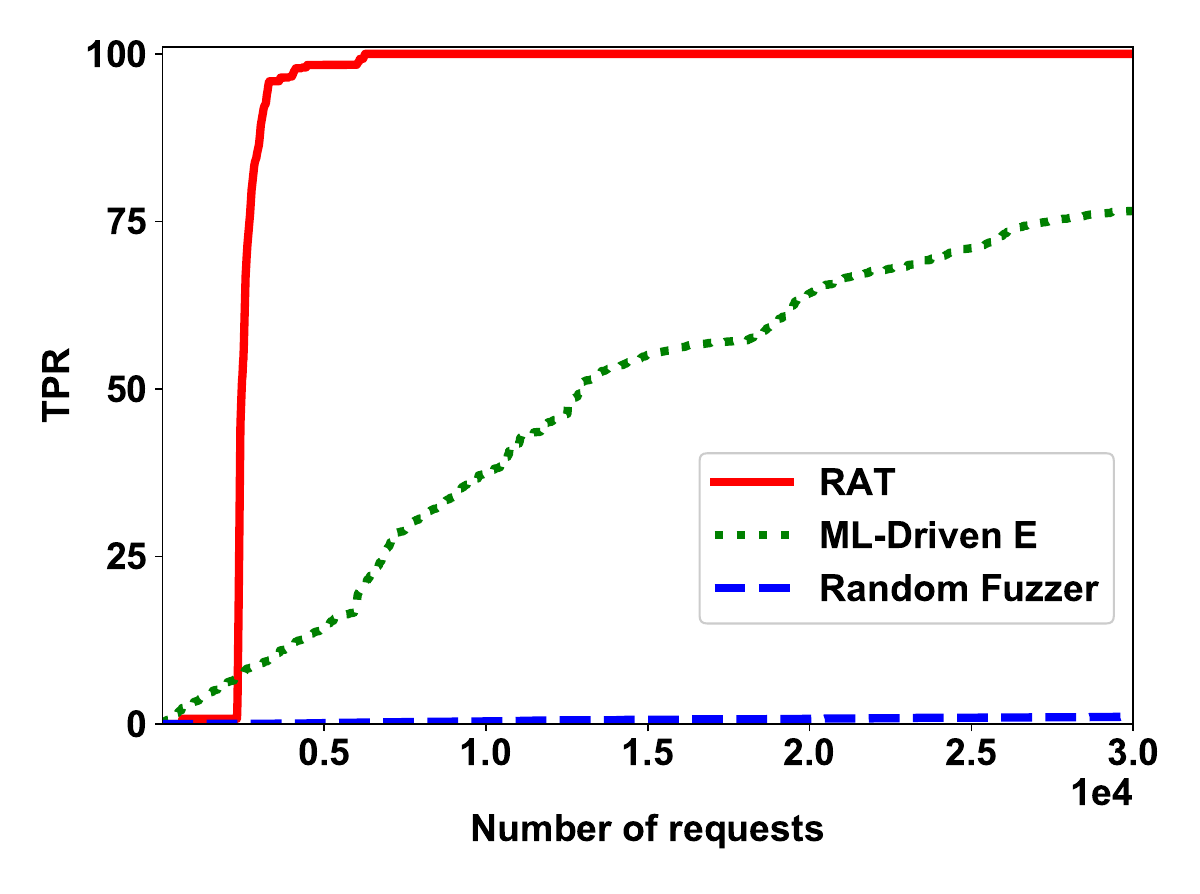}%
}
\caption{Average true positive rate in testing the open-source WAFs for SQLi vulnerabilities.}
\label{fig:compare_with_ml}
\vspace{-2em}
\end{figure}

To compare \textit{RAT} with \textit{Ml-Driven E} and the \textit{Random Fuzzer}, since the objective of \textit{Ml-Driven E} is to discover the highest possible number of SQLi vulnerabilities, we measured TPR over 30,000 requests for each approach. We then assess the results to compare their effectiveness.

\textit{ART4SQLi} and \textit{XSSART} tend to find the very first bypassing payload with the lowest number of requests. Therefore, we measured FPs before finding the first bypassing payload to compare \textit{RAT} with these techniques. For this purpose in each episode, we selected a random cluster and applied the \textit{AdaptiveSearch} to the chosen cluster for one round.

We applied all techniques to open-source WAFs, and then we calculated the average TPR for all parameters (Section~\ref{procedure}). Comparative tests between \textit{RAT}, \textit{Ml-Driven E} and the \textit{Random Fuzzer} are repeated 30 times as these experiments are time-consuming, and increasing the number of tests was not feasible with our resources. On the other hand, comparative tests between \textit{RAT}, \textit{ART4SQLi} and \textit{XSSART} are not time-consuming. Moreover, since \textit{ART4SQLi} and \textit{XSSART} are at most 27\% more efficient than random testing, their tests require more repetition to report reliable results. Therefore, we repeated the comparative tests between \textit{RAT}, \textit{ART4SQLi} and \textit{XSSART} 100 times.

To statistically compare different methods, we used Wilcoxon rank-sum test with the significance level of $\alpha=0.05$. Since this test is non-parametric, it does not require the samples to be normally distributed. To perform the Wilcoxon test, we collected all test results for each repetition and then grouped them by the type of attack (e.g., SQLi or XSS). Thus, we report the Wilcoxon results for each type of attack individually.

\begin{figure}[!t]
\centering
  \includegraphics[width=0.7\linewidth]{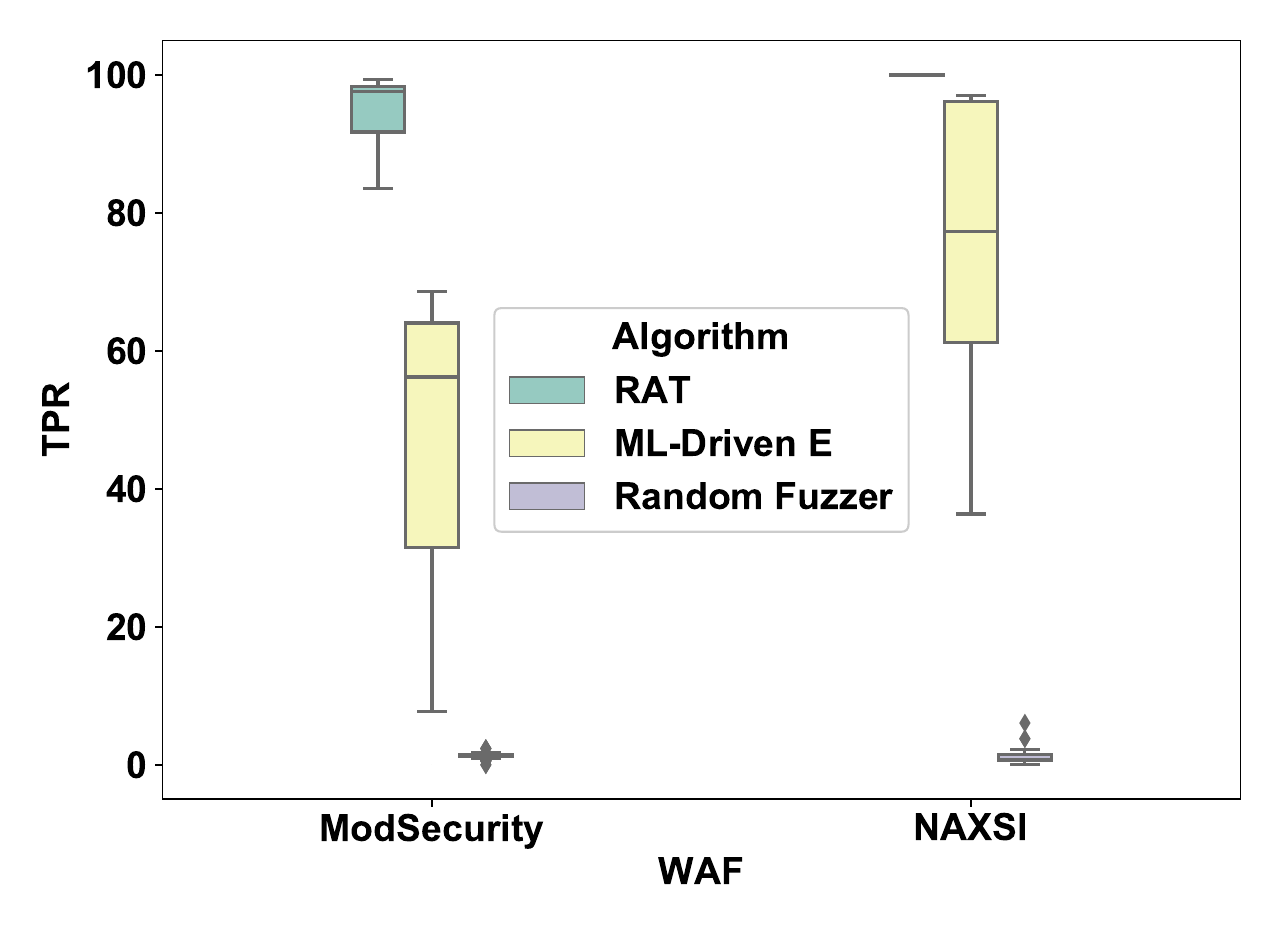}
  \caption{Boxplots of the average true positive rate in testing the open-source WAFs for SQLi vulnerabilities.}
  \label{fig:compare_with_ml_box}
  \vspace{-0.5em}
\end{figure}

\figurename~\ref{fig:compare_with_ml} depicts the result of the comparison between \textit{RAT}, \textit{Ml-Driven E} and the \textit{Random Fuzzer}, and \figurename~\ref{fig:compare_with_ml_box} illustrates the same result in the form of boxplots to visualize statistical variation. As shown in \figurename~\ref{fig:compare_with_ml}, within 30,000 requests, the \textit{RAT} could achieve an average of 95.37\% and 100\% TPR in testing \textit{ModSecurity} and \textit{NAXSI}, respectively. Moreover, in the worst case, It could find 83.53\% of bypassing payloads (\figurename~\ref{fig:compare_with_ml_box}). In comparison, in the best case, \textit{Ml-Driven E} could find 48.51\% and 76.56\%  of bypassing payloads in testing \textit{ModSecurity} and \textit{NAXSI}, respectively. The observations reveal that \textit{RAT} and \textit{Ml-Driven E} can find bypassing attacks, whereas \textit{Random Fuzzer} failed due to the rarity of bypassing attacks. Moreover, the results clearly show that the \textit{RAT} has a lower false-positive rate and significantly outperforms counterparts.

\begin{table}[!t]
\caption{Result of the Wilcoxon test in testing SQLi vulnerabilities of ModSecurity and NAXSI.}
\label{tab:wilcoxon_sqli_local}
\begin{adjustbox}{center}
\renewcommand{\arraystretch}{1.5}
\begin{tabular}{|c|c|c|c|}
\hline
Requests & Random Fuzzer & \begin{tabular}[c]{@{}c@{}}Ml-Driven E\\ $(p<.001)$\end{tabular} & \begin{tabular}[c]{@{}c@{}}RAT\\ $(p<.001)$\end{tabular}    \\ \hline
$10,000$             & ---           & Random Fuzzer                                                            & \begin{tabular}[c]{@{}c@{}}Ml-Driven E\\ Random Fuzzer\end{tabular} \\ \hline
$20,000$             & ---           & Random Fuzzer                                                            & \begin{tabular}[c]{@{}c@{}}Ml-Driven E\\ Random Fuzzer\end{tabular} \\ \hline
$30,000$             & ---           & Random Fuzzer                                                            & \begin{tabular}[c]{@{}c@{}}Ml-Driven E\\ Random Fuzzer\end{tabular} \\ \hline
\end{tabular}
\end{adjustbox}
\vspace{-0.5em}
\end{table}

Table~\ref{tab:wilcoxon_sqli_local} shows how each approach outperforms others after each $10,000$ requests. We measured TPR after each $10,000$ requests for each repetition to create this table and then used the Wilcoxon test to compare TPRs of different methods statistically. In Table~\ref{tab:wilcoxon_sqli_local}, each cell represents the approaches outperformed by the approach matching the corresponding column when the $p-values<.001$.

\begin{table}[!t]
\caption{Number of false positives before finding the first SQLi bypassing payload.}
\label{tab:rat_art4sqli}
\begin{adjustbox}{center}
\renewcommand{\arraystretch}{1.5}
\begin{tabular}{|c|c|c|c|c|}
\hline
\multirow{2}{*}{Method} & \multicolumn{2}{c|}{GET Parameter} & \multicolumn{2}{c|}{Cookie} \\ \cline{2-5} 
                        & ModSecurity         & NAXSI        & ModSecurity     & NAXSI     \\ \hline
RAT                     & 142.4               & 209.3        & 179.33          & 210.78    \\ \hline
ART4SQLi                & ---                 & ---          & 465.02          & ---       \\ \hline
Random                  & ---                 & ---          & ---             & ---       \\ \hline
\end{tabular}
\end{adjustbox}
\vspace{-2em}
\end{table}

Table~\ref{tab:rat_art4sqli} shows the result of the comparison between \textit{RAT}, \textit{ART4SQLi} and random technique. The observations show that \textit{RAT} could find the first bypassing with the reasonable number of false positives, whereas \textit{ART4SQLi} failed to find a bypassing payload within the limited number of attempts except for the cookie parameter of \textit{ModSecurity} in which \textit{RAT} was 61.43\% faster than \textit{ART4SQLi} on average. We also compared FPs in testing cookie parameter of \textit{ModSecurity} using the Wilcoxon test. As a result, \textit{RAT} could outperform \textit{ART4SQLi} with the $p-values<.001$.

Furthermore, we performed the same experiment on \textit{RAT} and \textit{XSSART} using the XSS dataset. We observed that for ModSecurity, \textit{RAT} could find the first bypassing payload after 581.45 unsuccessful attempts and failed in finding the first bypassing payload within the limited number of attempts in testing NAXSI. In comparison, \textit{XSSART} failed in both situations as well as random techniques.

\begin{figure}[t!]

\subfloat[ModSecurity]{%
  \includegraphics[width=0.5\columnwidth]{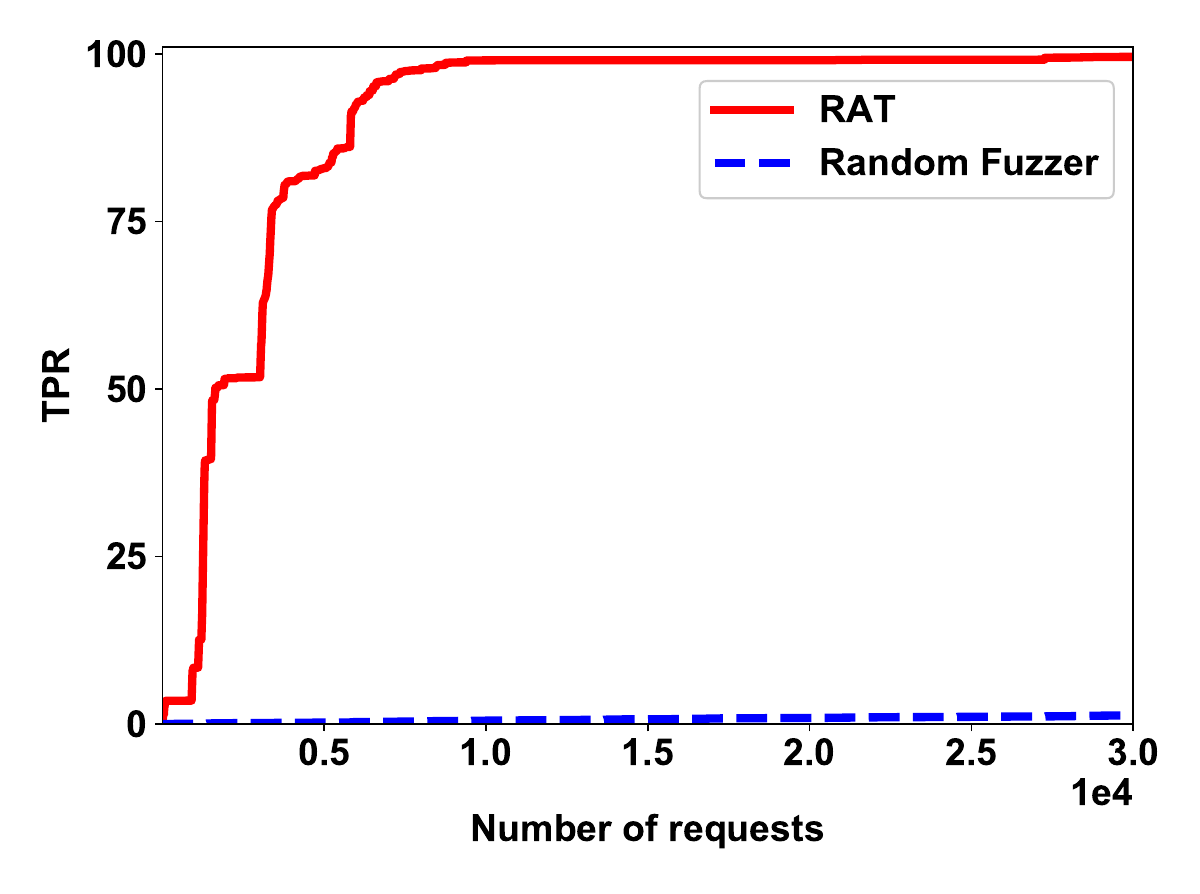}%
}
\hfill
\subfloat[NAXSI]{%
  \includegraphics[width=0.5\columnwidth]{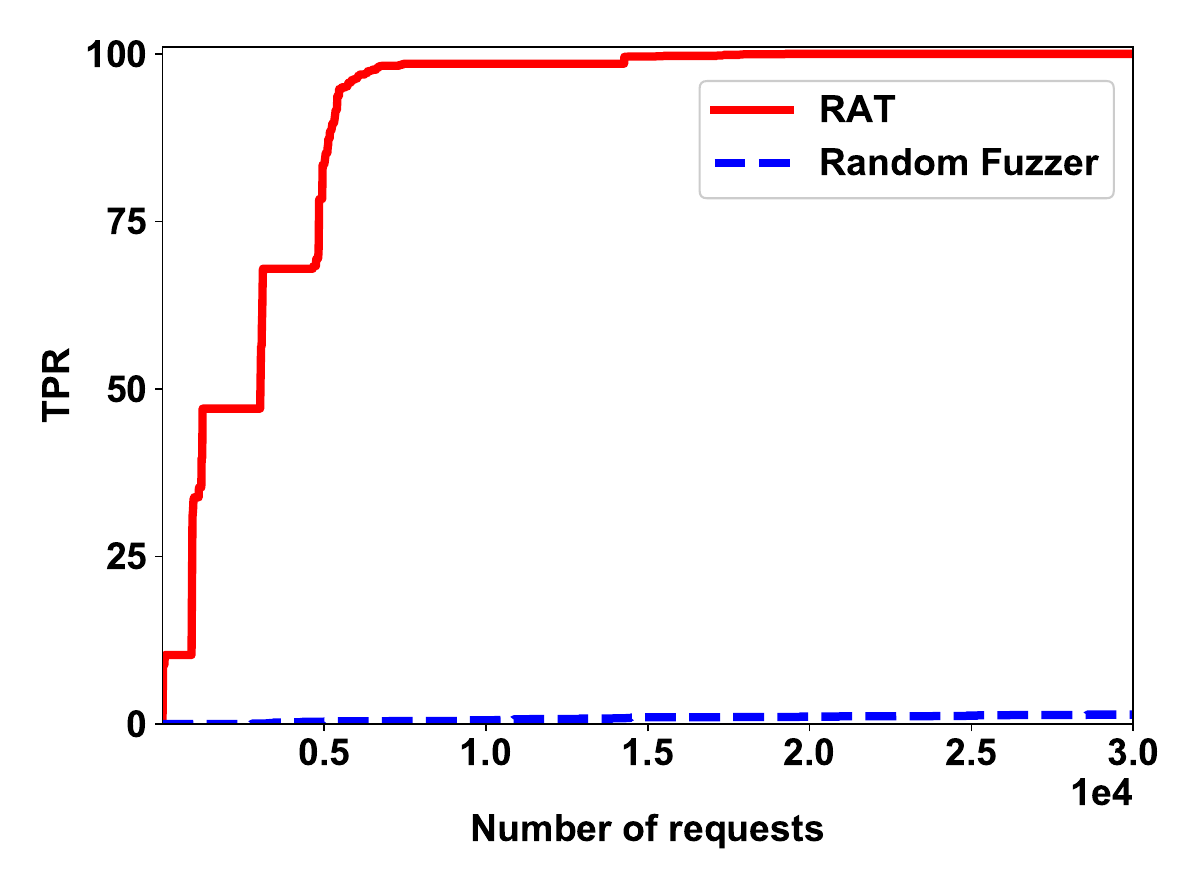}%
}

\caption{Average true positive rate in testing the open-source WAFs for XSS vulnerabilities.}
\label{fig:compare_with_rand_xss}
\vspace{-1em}
\end{figure}
\begin{figure}[!t]
\centering
  \includegraphics[width=0.7\linewidth]{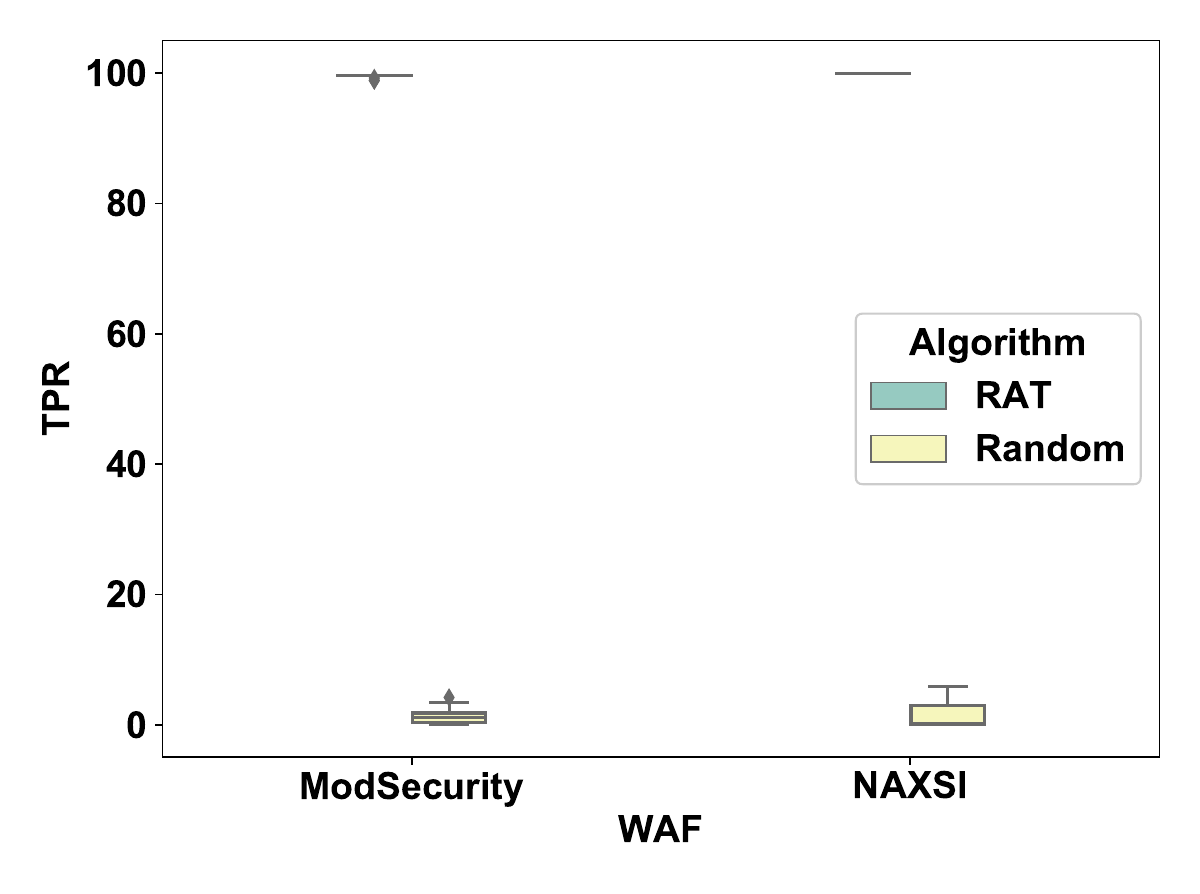}
  \caption{Boxplots of the average true positive rate in testing the open-source WAFs for XSS vulnerabilities.}
  \label{fig:compare_with_rand_box}
  \vspace{-2em}
\end{figure}

Finally, to evaluate the performance of \textit{RAT} at testing a different attack type, we applied \textit{RAT} to \textit{ModSecurity} and \textit{NAXSI} using the XSS dataset. \figurename~\ref{fig:compare_with_rand_xss} depicts the result of testing \textit{ModSecurity} and \textit{NAXSI} for XSS vulnerabilities using \textit{RAT} and the Random Fuzzer, and \figurename~\ref{fig:compare_with_rand_box} shows the statistical variation of the test results. According to \figurename~\ref{fig:compare_with_rand_xss}, \textit{RAT} could find 100\% of bypassing payloads before reaching 10,000 requests, whereas the \textit{Random Fuzzer} could only discover 1\% of bypassing payloads. Moreover, we compared these two methods using Wilcoxon test. The result verified that \textit{RAT} can clearly outperform \textit{Random Fuzzer} when the $p-values<.001$.

\vspace{-1em}
\subsection{\textit{Q4: Is the efficiency and effectiveness of RAT acceptable in practice?}}
To answer the final question, we applied \textit{RAT} and its counterparts to the custom-built WAF over the internet (30 repetitions for each test). To evaluate the effectiveness, we tested 30,000 payloads with each method and measured the number of uncovered bypassing payloads (TP). Since it is a real-world application, the brute-force attack was infeasible, and we could not measure the total number of bypassing attacks to calculate TPR. Therefore we only report TP.

\begin{figure}[t!]

\subfloat[Line Chart\label{fig:compare_with_ml_uni}]{%
  \includegraphics[width=0.5\columnwidth]{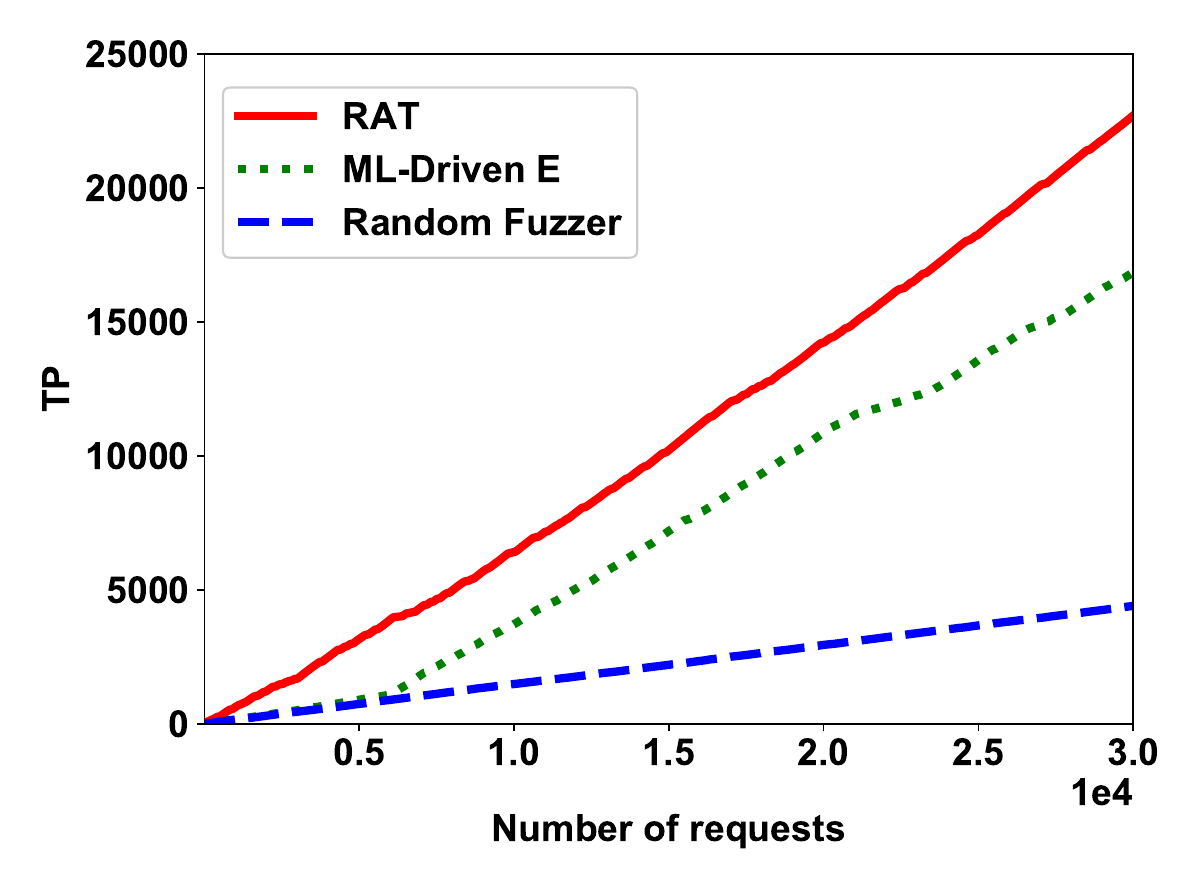}%
}
\hfill
\subfloat[Boxplots\label{fig:compare_with_ml_uni_box}]{%
  \includegraphics[width=0.5\columnwidth]{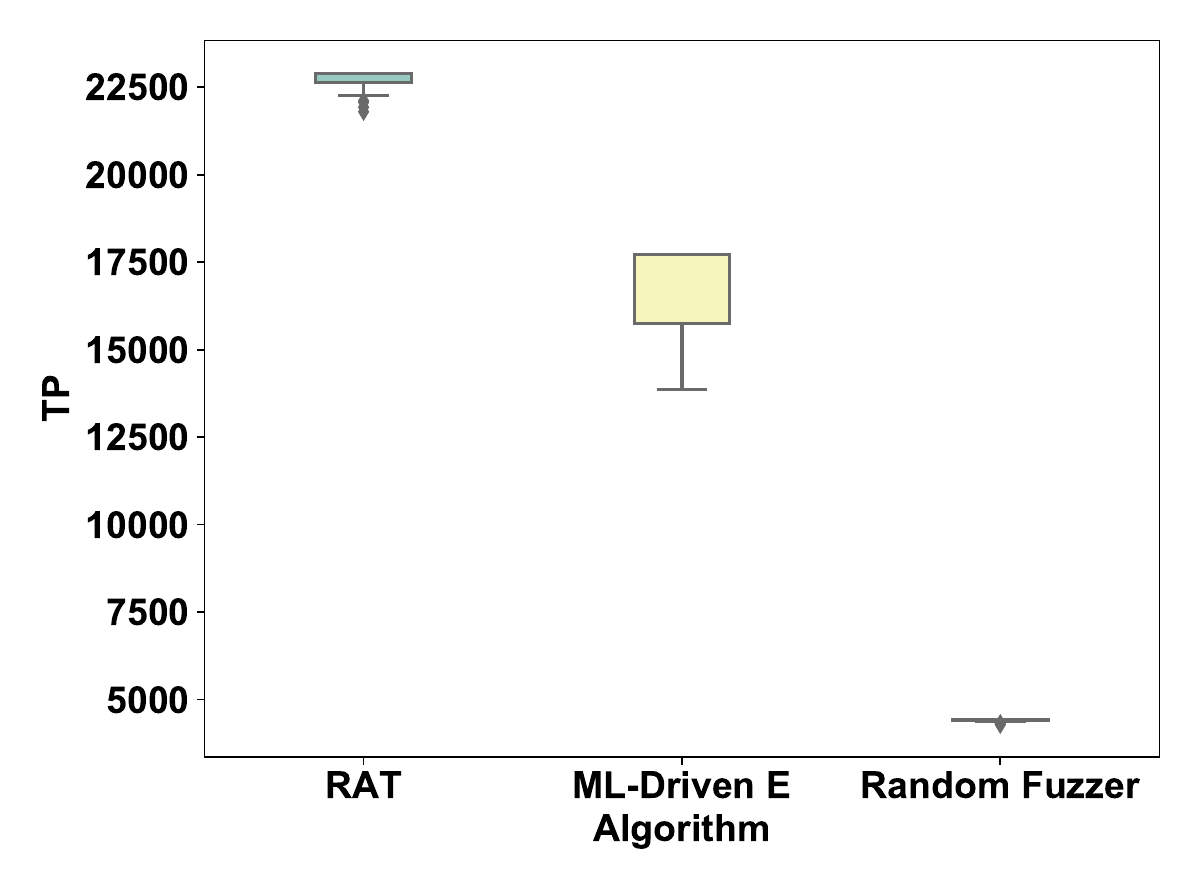}%
}

\caption{Average number of bypassing attacks in testing the custom-built WAF for SQLi vulnerabilities.}

\vspace{-1.5em}
\end{figure}

\begin{figure}[t!]

\subfloat[Line Chart\label{fig:compare_with_ran_uni}]{%
  \includegraphics[width=0.5\columnwidth]{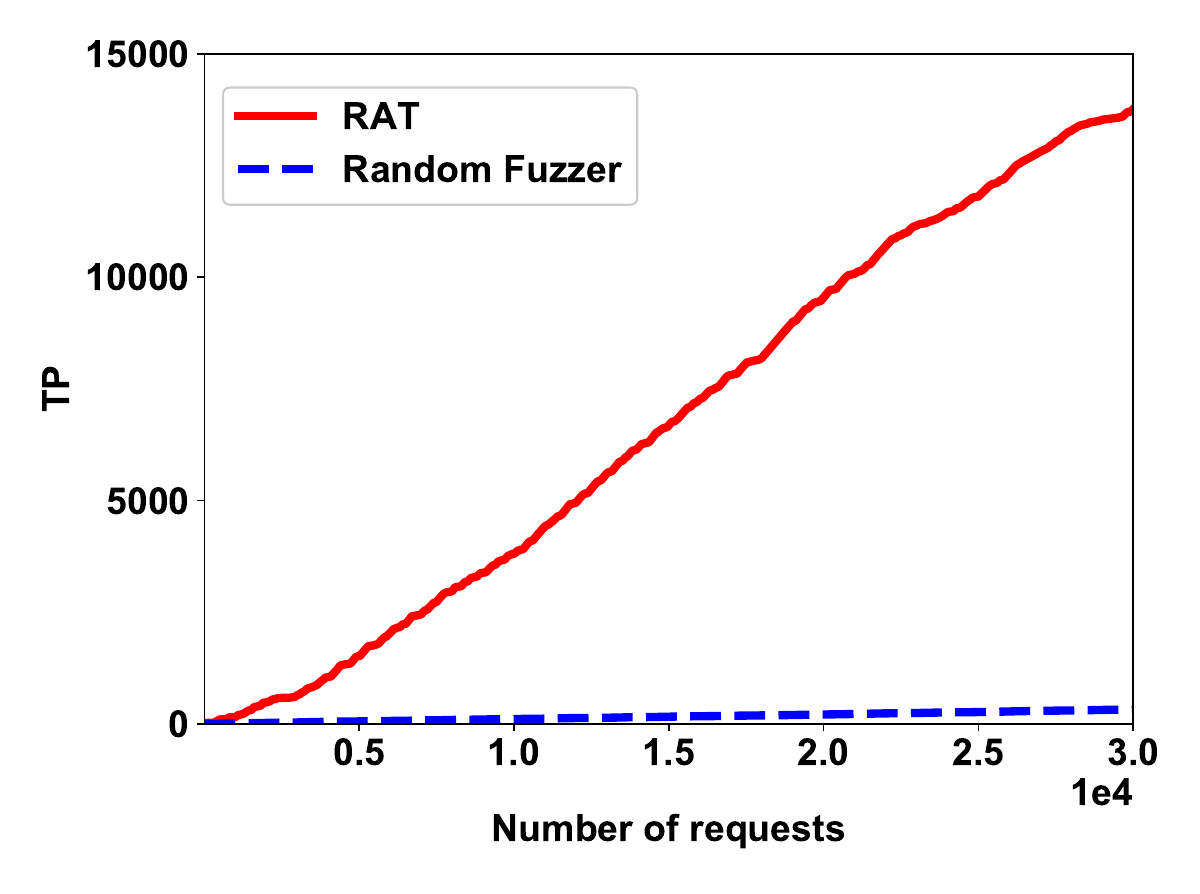}%
}
\hfill
\subfloat[Boxplots\label{fig:compare_with_rand_uni_box}]{%
  \includegraphics[width=0.5\columnwidth]{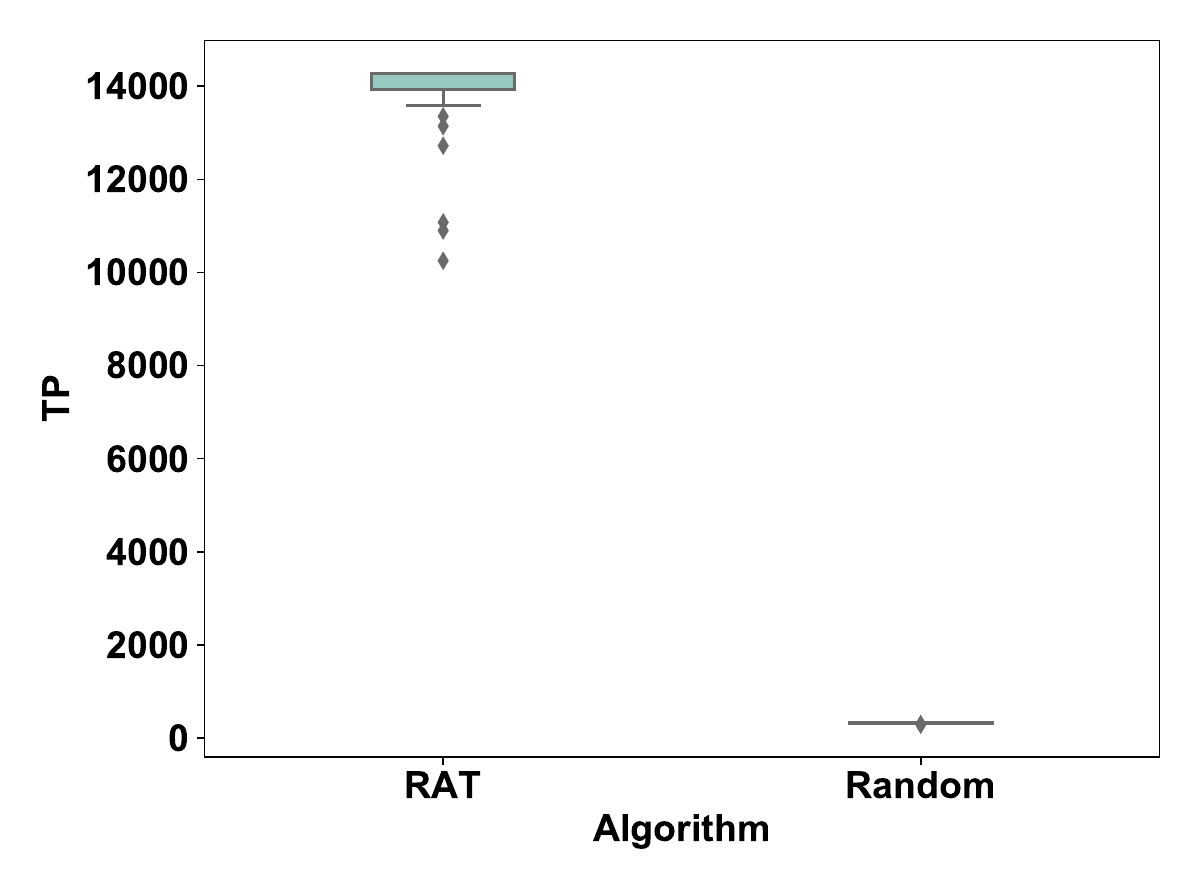}%
}

\caption{Average number of bypassing attacks in testing the custom-built WAF for XSS vulnerabilities.}

\vspace{-0.5em}
\end{figure}

\begin{table}[!t]
\caption{The result of comparison between \textit{RAT}, \textit{ART4SQLi}, and \textit{Random Fuzzer} in testing the custom-built WAF for SQLi vulnerabilities. In this table, the mean is the average FP before finding the first bypassing payload.}
\label{tab:rat_art4sqli_internet}
\begin{adjustbox}{center}
\renewcommand{\arraystretch}{1.5}
\begin{tabular}{c|c|c|c|}
\cline{2-4}
                                      & Random                  & ART4SQLi               & RAT  \\ \hline
\multicolumn{1}{|c|}{Mean}            & 4.93                    & 4.64                   & 7.57 \\ \hline
\multicolumn{1}{|c|}{Std}             & 5.96                    & 4.63                   & 7.21 \\ \hline
\multicolumn{1}{|c|}{Wilcoxon result} & RAT ($p<0.001$) & RAT ($p<0.01$) & ---  \\ \hline
\end{tabular}
\end{adjustbox}
\vspace{-0em}
\end{table}

\begin{table}[!t]
\caption{The result of comparison between \textit{RAT}, \textit{XSSART}, and \textit{Random Fuzzer} in testing the custom-built WAF for XSS vulnerabilities. In this table, the mean is the average FP before finding the first bypassing payload.}
\label{tab:rat_xssart_internet}
\begin{adjustbox}{center}
\renewcommand{\arraystretch}{1.5}
\begin{tabular}{c|c|c|c|}
\cline{2-4}
                                      & Random & XSSART & \begin{tabular}[c]{@{}c@{}}RAT\\ ($p<0.001$)\end{tabular} \\ \hline
\multicolumn{1}{|c|}{Mean}            & 92.95  & 107.61 & 37.77                                                             \\ \hline
\multicolumn{1}{|c|}{Std}             & 92.16  & 116.63 & 15.65                                                             \\ \hline
\multicolumn{1}{|c|}{Wilcoxon result} & ---    & ---    & \begin{tabular}[c]{@{}c@{}}Random\\ XSSART\end{tabular}           \\ \hline
\end{tabular}
\end{adjustbox}
\vspace{-2em}
\end{table}
\figurename~\ref{fig:compare_with_ml_uni} shows the result of applying the three methods to the custom-built WAF for SQLi discovery, and \figurename~\ref{fig:compare_with_ml_uni_box} shows the same result in the form of box-plots for better statistical visualization. It shows that the \textit{RAT} can uncover a significant number of bypassing payloads within a reasonable number of requests. On average, \textit{RAT} sent 2.83 requests for each bypassing payload, whereas this number for \textit{Ml-Driven E} is 4.06 and 13.56 for the Random method.

To evaluate the efficiency of \textit{RAT}, we measured the average TSR for \textit{RAT} and \textit{Ml-Driven E}. The average TSR value for \textit{RAT} was 0.74 seconds, and for \textit{Ml-Driven E}, this value was 0.80 seconds. The TSR values for both methods were almost the same, and it is a reasonable value in practice.

We also conducted the same experiment for the XSS attack (see \figurename~\ref{fig:compare_with_ran_uni} and \ref{fig:compare_with_rand_uni_box}). The results are as follows.

\begin{enumerate}
\item RAT sent 4.44 requests on average, whereas the Random technique sent 184.11 requests.
\item The average TSR value for \textit{RAT} in this experiment was 0.39 seconds, which was lower than the TSR value of the SQLi test due to the fewer samples.
\end{enumerate}

We conducted the Wilcoxon test to compare TPs of \textit{RAT} with \textit{Ml-Driven E} and \textit{Random Fuzzer}. The results of testing SQLi were the same as in Table 9, and in testing XSS, \textit{RAT} could outperform \textit{Random Fuzzer} when the $p-values<.001$.

We also repeated our comparative experiments between \textit{RAT}, \textit{ART4SQLi}, and \textit{XSSART} for the custom-built WAF to compare the performance of these approaches in practice. The results are shown in Tables~\ref{tab:rat_art4sqli_internet} and~\ref{tab:rat_xssart_internet}. According to Table~\ref{tab:rat_art4sqli_internet}, Random strategy and \textit{ART4SQLi} could discover the first bypassing payload with fewer attempts than \textit{RAT}. Wilcoxon results also verify that Random strategy and \textit{ART4SQLi} could outperform \textit{RAT}. However, neither \textit{ART4SQLi} nor Random strategy could outperform the other one. The possible reason for the obtained results can be the massive number of SQLi vulnerabilities of the custom-built WAF. Despite the results shown in Table~\ref{tab:rat_art4sqli_internet}, Table~\ref{tab:rat_xssart_internet} shows that \textit{RAT} could significantly perform better than \textit{XSSART} and the Random strategy in discovering the first XSS bypassing payload.

In conclusion, our answer to the Q4 is that yes, \textit{RAT} is efficient and effective in practice.
\vspace{-1em}
\section{Conclusion} \label{conclusion}
In this paper, we proposed the \textit{RAT}, a search-based technique that combines a reinforcement learning algorithm with an innovative adaptive search method. \textit{RAT} automatically extracts patterns from attack payloads. It then clusters similar payloads together and discovers the clusters which contain bypassing payloads using a reinforcement learning technique. Finally, \textit{RAT} ranks test candidates and selects the payload with the highest probability of bypassing the WAF.

Empirical results suggested that considering the sequence of tokens rather than single literals and clustering payloads improves the performance of discovering effective payloads. Comparative experiments, moreover, showed that \textit{RAT} significantly performs better than the state-of-the-art algorithms. Finally, the result of applying \textit{RAT} to a real-world WAF demonstrated that \textit{RAT} is efficient and effective in practice. However, \textit{RAT} is highly dependent on the dataset, and the comprehensiveness of the dataset directly affects the \textit{RAT}'s performance. Furthermore, \textit{RAT} can only test rule-based WAFs, and it cannot be used alone to test Ml-based WAFs. Nevertheless, \textit{RAT} is capable of being combined with generative adversarial techniques (e.g., \textit{WAF-A-MoLE} \citep{demetrio2020waf}) to reduce its dependency on datasets and test Ml-based WAFs.

In our future studies, we will work on combining \textit{RAT} with Generative Adversarial Networks to propose an adversarial test strategy as well as a solution for altering WAFs. We will also focus on building comprehensive datasets with the least number of samples, and we will investigate different clustering strategies to improve \textit{RAT}'s efficiency.
\vspace{-1em}
\section{acknowledgement}
The authors would like to acknowledge the financial support of Information and Communication Technology Park for this project under grant number 16-99-01-000040.

\vspace{-1em}
\ifCLASSOPTIONcaptionsoff
  \newpage
\fi



%
\bibliographystyle{IEEEtranN}
\bibliography{References.bib}


%
%

\vskip -2\baselineskip plus -1fil
\begin{IEEEbiography}[{\includegraphics[width=1in,height=1.25in,clip,keepaspectratio]{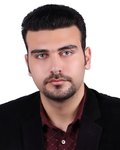}}]{Mohammadhossein Amouei}
is an MSc student in the Faculty of Computer Engineering at the Shahrood University of Technology, Shahrood, Iran. His research interests include artificial intelligence and computer security.
\end{IEEEbiography}
\vskip -2\baselineskip plus -1fil
\begin{IEEEbiography}[{\includegraphics[width=1in,height=1.25in,clip,keepaspectratio]{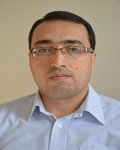}}]{Mohsen Rezvani}
received the PhD degree in computer science from the University of New South Wales, Australia. He is a faculty member in the Faculty of Computer Engineering, Shahrood University of Technology, Iran. His research focuses on computer security, privacy, trust and reputation systems.
\end{IEEEbiography}
\vskip -2\baselineskip plus -1fil

\begin{IEEEbiography}[{\includegraphics[width=1in,height=1.25in,clip,keepaspectratio]{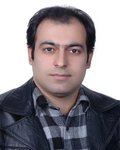}}]{Mansoor Fateh}
received the M.S. degree in Biomedical Engineering from Tarbiat Modares University, Tehran, Iran, and the PhD from Tarbiat Modares University, Tehran, Iran. He is a faculty member in the Faculty of Computer Engineering, Shahrood University of Technology, Iran. His research interests include machine learning and image processing.
\end{IEEEbiography}





\end{document}